\PassOptionsToPackage{table,dvipsnames}{xcolor}
\documentclass[acmtog, authorversion, screen]{acmart}

\usepackage{booktabs} 

\setcitestyle{square}

\acmSubmissionID{986}

\usepackage{amsbsy}
\usepackage[mathscr]{euscript}
\usepackage{amsmath}
\DeclareMathOperator*{\argmax}{arg\,max}
\usepackage{stackrel}
\usepackage{nicefrac}

\usepackage{pifont}
\usepackage{booktabs}
\usepackage{colortbl}
\usepackage{array}
\usepackage{ragged2e}
\usepackage{tabularx}

\usepackage{makecell}
\usepackage{multirow}
\usepackage{marvosym}

\usepackage{graphicx}
\usepackage{siunitx}
\usepackage{algpseudocode}
\usepackage[ruled]{algorithm2e} 

\SetAlFnt{\small}
\SetAlCapFnt{\small}
\SetAlCapNameFnt{\small}
\SetAlCapHSkip{0pt}
\IncMargin{-\parindent}

\usepackage{dblfloatfix}
\usepackage{float}

\usepackage{array}

\newcolumntype{L}[1]{>{\raggedright\let\newline\\\arraybackslash\hspace{0pt}}m{#1}}
\newcolumntype{C}[1]{>{\centering\let\newline\\\arraybackslash\hspace{0pt}}m{#1}}

\usepackage{siunitx}
\usepackage{subcaption}

\usepackage{overpic}

\usepackage[version=4]{mhchem}

\setcopyright{cc}
\setcctype{none}
\acmJournal{TOG}
\acmYear{2025} \acmVolume{44} \acmNumber{4} \acmArticle{} \acmMonth{8} \acmPrice{}\acmDOI{10.1145/3731200}

\newcommand\rev[1]{\textcolor{black}{#1}}
\newcommand\supp{\textit{Supplementary Material}}

\makeatletter
\DeclareRobustCommand\onedot{\futurelet\@let@token\@onedot}
\def\@onedot{\ifx\@let@token.\else.\null\fi\xspace}

\def\eg{\emph{e.g}\onedot} 
\def\ie{\emph{i.e}\onedot}

\begin{document}

\title{Collaborative On-Sensor Array Cameras}

\author{Jipeng Sun}
\email{jipeng.sun@princeton.edu}
\affiliation{%
  \institution{Princeton University}
  \country{USA}
}

\author{Kaixuan Wei}
\email{kaixuan.wei@kaust.edu.sa}
\affiliation{%
  \institution{KAUST}
  \country{Saudi Arabia}
}

\author{Thomas Eboli}
\email{thomas.eboli12@gmail.com}
\affiliation{%
  \institution{Université Paris-Saclay}
  \country{France}
}

\author{Congli Wang}
\email{congli.wang@princeton.edu}

\author{Cheng Zheng}
\email{chengzh@princeton.edu}

\affiliation{%
  \institution{Princeton University}
  \country{USA}
}

\author{Zhihao Zhou}
\email{zzhou99@uw.edu}

\author{Arka Majumdar}
\email{arka@uw.edu}
\affiliation{%
  \institution{University of Washington}
  \country{USA}
  }

\author{Wolfgang Heidrich}
\email{wolfgang.heidrich@kaust.edu.sa}
\affiliation{%
  \institution{KAUST}
  \country{Saudi Arabia}
  }

\author{Felix Heide}
\email{fheide@princeton.edu}
\affiliation{%
  \institution{Princeton University}
  \country{USA}
  }

\authorsaddresses{%
Authors’ addresses: Jipeng Sun, jipeng.sun@princeton.edu, Princeton University, USA; %
\Letter Kaixuan Wei, kaixuan.wei@kaust.edu.sa, KAUST, Saudi Arabia; %
Thomas Eboli, thomas.eboli12@gmail.com, Université Paris-Saclay, France; %
Congli Wang, congli.wang@princeton.edu, Princeton University, USA; %
Cheng Zheng, chengzh@princeton.edu, Princeton University, USA; %
Zhihao Zhou, zzhou99@uw.edu, University of Washington, USA; %
Arka Majumdar, arka@uw.edu, University of Washington, USA; %
Wolfgang Heidrich, wolfgang.heidrich@kaust.edu.sa, KAUST, Saudi Arabia; %
Felix Heide, fheide@princeton.edu, Princeton University, USA. %
}

\begin{abstract}
Modern nanofabrication techniques have enabled us to manipulate the wavefront of light with sub-wavelength-scale structures, offering the potential to replace bulky refractive surfaces in conventional optics with ultrathin metasurfaces. In theory, arrays of nanoposts provide unprecedented control over manipulating the wavefront in terms of phase, polarization, and amplitude at the nanometer resolution. A line of recent work successfully investigates flat computational cameras that replace compound lenses with a single metalens or an array of metasurfaces a few millimeters from the sensor. However, due to the inherent wavelength dependence of metalenses, in practice, these cameras do not match their refractive counterparts in image quality for broadband imaging, and may even suffer from hallucinations when relying on generative reconstruction methods.

In this work, we investigate a collaborative array of metasurface elements that are jointly learned to perform broadband imaging. To this end, we learn a  nanophotonics array \rev{with 100-million nanoposts} that is end-to-end jointly optimized over the full visible spectrum---a design task that existing inverse design methods or learning approaches cannot support due to memory and compute limitations. We introduce a distributed meta-optics learning method to tackle this challenge. This allows us to optimize a large parameter array along with a learned metaatom proxy and a non-generative reconstruction method that is parallax-aware and noise-aware. The proposed camera performs favorably in simulation and in all experimental tests irrespective of the scene illumination spectrum.
\end{abstract}

\begin{teaserfigure}  
\begin{center}
    \hspace*{-8pt}\includegraphics[width=1.03\linewidth]{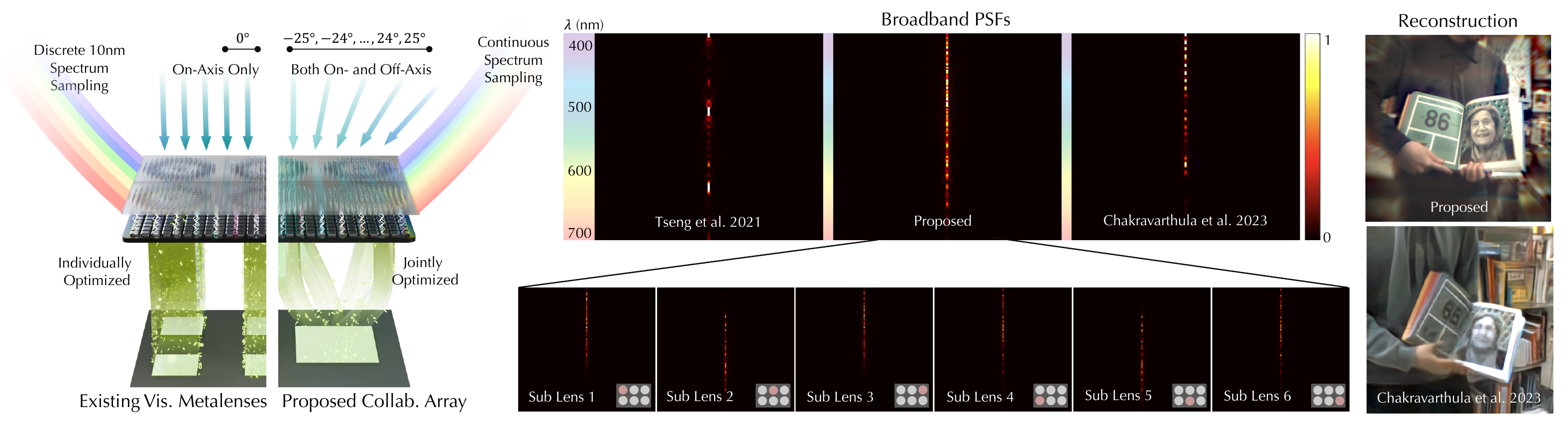}\vspace*{-8pt}
    \caption{We designed a collaborative metasurface array imager of $2\times3$ metalenses (left) positioned 3.6 mm from the sensor plane that enables broadband focusing (see 1-D PSF sections alongside spectrum in the middle) across the full \SI{50}{\degree}$\times$\SI{50}{\degree} field of view, including both axial and off-axis incident angles. The design collaboratively optimizes 100 million nanophotonic parameters through a neural structure-to-phase proxy and a distributed meta-optics training framework incorporating a parallax-aware spatially-varying method for accurate, hallucination-free image reconstruction (right). In reconstructing a book containing `86' and portrait, our collaborative metalens array accurately captures both elements, while existing flat metalens arrays~\cite{chakravarthula2023thin} produces hallucinated reconstructions that more closely resemble `65'.}
    \label{fig:teaser}
\end{center}
\end{teaserfigure}


\maketitle


\section{introduction}
\label{sec:intro}

The aperture of an optical system fundamentally limits its optical resolution. As such, large aperture optics are essential for applications across diverse domains, including photography, robotics, remote sensing, mobile smartphones, health, and scientific imaging. Large optics generally result in large weight; for example, the weight of the lenses in the Hubble space telescope (aperture of $\sim$~\SI{2}{\metre}) is on the order of $\sim$~\SI{1000}{\kilo\gram}. Enabled by modern nanofabrication techniques, metaoptics~\cite{khorasaninejad2017metalenses} have been proposed as a potential alternative to refractive surfaces. Metaoptics consist of quasi-periodic arrays of sub-wavelength scatterers that offer access to a design space that is six orders of magnitude larger than conventional spherical optics. These degrees of freedom offer unprecedented control over the modulation of the incoming wavefront, with the potential to drastically reduce the size and weight of conventional large-aperture optics.


However, traversing this larger design space to find large-aperture optics that offer substantially smaller footprint without compromising image quality has proven challenging, even with recent learned~\cite{neural_nanooptics, chakravarthula2023thin, froch2025beating} design approaches. This is because diffractive optical elements (DOEs) and metasurfaces are fundamentally wavelength-dependent chromatic optics, which makes it challenging to minimize chromatic aberration while achieving high resolution and large field of view~(FOV). Although techniques like dispersion engineering have been explored to address chromatic aberrations~\cite{chen2018broadband,Wang2018ABA}, the limited optical bandwidth and the aperture size lead to poor imaging performance under broadband illumination~\cite{Presutti2020FocusingLimits}. A recent line of work has explored learning methods for optimizing metalenses or arrays of metasurfaces using proxy forward models~\cite{neural_nanooptics,chakravarthula2023thin}. To date, the most successful meta-optics imager for broadband imaging is inverse designed via a differentiable broadband image formation model~\cite{chakravarthula2023thin}. 
Despite the improvements in image quality, the design limitations of the heuristic wedge phase superimposition results in compromised color fidelity and severe distortions in spatial frequencies, requiring a generative model for image recovery with the risk of hallucinations. As a direct result of the large size of the parameter space, only the center lens profile could be learned while the peripheral lenses were directly made by superimposing the center lens profile with a certain wedge phase to heuristically expand FOV. Consequently, measurements from this existing camera (especially the peripheral lenses) suffer from severe degradations in spatial frequencies and color fidelity. Although significant strides in meta-optics imaging have been made in both optics and computational imaging communities, the image quality of ultra-thin metalens cameras is still far from being competitive with traditional refractive optics.

In this work, we propose a flat camera composed of an array of metalenses which are jointly and collaboratively designed for broadband imaging. To achieve this, we learn a nanophotonic array with  \rev{100-million-nanoposts} that is end-to-end optimized over the full visible spectrum – a design task existing inverse design methods or learning approaches cannot support due to memory and compute limitations.  Our design is motivated by the observations that 1) a single metalens can possess favorable focusing properties within a narrow band, and 2) a sequence of measurements that are optically encoded in distinct ways can be collaboratively merged to generate high-quality results. We find that the collaborative array features per-lens optical encodings, each of which is tailored to distinct yet complementary segments of the color spectrum. To design the array we develop a distributed, large-scale meta-optics optimization framework.
We learn the array optics with a novel learned nano-structure-to-broadband-phase neural proxy. To recover images, we introduce a collaborative parallax-aware joint reconstruction model that does not rely on generative priors. We fabricate the designed metasurface array and build a prototype camera system to experimentally validate the proposed design method. We confirm that the collaborative metalens array camera consistently outperforms the existing meta-optics in terms of image quality across the entire visible spectrum, while maintaining a flat optical system on the sensor cover glass.

\medskip
Our contributions are as follows:
\begin{itemize}
    \item We introduce a collaborative on-sensor array camera, where each lens learns to capture complementary portions of the spectrum. When combined, these measurements enable high-fidelity reconstruction across the full visible spectrum range.
    \item We propose a learned nano-structure-to-broadband-phase neural proxy for accurate forward simulation, and a distributed large-scale meta-optics optimization method.
    \item \rev{We devise a parallax‐aware, spatially‐variant reconstruction algorithm tailored to our jointly designed lens array, deliberately avoiding reliance on generative priors.}
    \item We evaluate our approach in simulations and in real indoor and outdoor scenes with a camera prototype, confirming favorable broadband imaging performance.
\end{itemize}

\paragraph{Limitations}

\rev{Compared to conventional cameras that employ larger refractive lens assemblies to record high-quality images without computationally heavy reconstruction, the proposed flat-camera does require computational recovery to produce an image. Moreover, as a research prototype rather than a commercial product, the current camera prototype retains the thick sensor cover glass that limits inner-baffle height, precludes further focal-length reduction, and mandates a non-contiguous sublens arrangement to avoid crosstalk. These compromises enlarge the required sensor size and limit the overall light-collection efficiency.}



\section{Related Work}
\label{sec:related}
\paragraph{Flat Computational Cameras.}
Designing a flat camera system with high imaging quality and low cost has been an open challenge that a large body of work in computational imaging addresses. As optical aberrations inherently limit the quality of flat optical systems, existing approaches share the same philosophy that trades off camera form factor with image quality and computational complexity~\cite{heide2013high,venkataraman2013picam,heide2016encoded,Peng2015ComputationalIU,Peng2016DiffractiveAchromat,asif2016flatcam,LearnedLargeFovImaging2019,khan2020flatnet,antipa2018diffusercam,Boominathan2016LenslessImaging,boominathan2020phlatcam}. Early attempts at flat computational camera designs compress the compound camera optics into a single refractive element (\eg, plano-convex or biconvex lenses~\cite{kingslake2009lens}), and they correct the geometric and chromatic aberrations post-capture using variational optimization~\cite{heide2013high,schuler2011non}. To further shrink the form factor, researchers have explored DOEs building atop diffraction manipulation of light beyond refraction, including single DOE elements (\eg, Fresnel lens~\cite{heide2016encoded,Peng2015ComputationalIU,Peng2016DiffractiveAchromat}), lensless coded apertures~\cite{asif2016flatcam,boominathan2020phlatcam}, diffusers~\cite{kuo2017diffusercam,monakhova2020spectraldiffuser}, and Fresnel zone apertures~\cite{wu2020single}. 
The compound cameras can achieve aberration-corrected high-quality imaging under broadband illumination, but still entails a long back-focal distance of more than \SI{10}{\milli\metre} which has made \textit{truly compact} camera design challenging. The DOE cameras, albeit allowing for ultra-thin cameras of a few millimeters in height, achieve moderate imaging quality due to the highly ill-posed nature of image reconstruction from coded measurements with global support. As such, unfortunately, no existing methods achieve flat form factors with high imaging quality.

\paragraph{Flat Metasurface Optics.}
Unlike conventional optical elements that rely on gradual phase shifts accumulated during light propagation to shape light beams, metasurfaces -- imparting abrupt phase changes with subwavelength nanostructures -- have recently been investigated for imaging~\cite{Yu2011LightPW,Yu2014FlatOW,Kildishev2013PlanarPW,Khorasaninejad2016MetalensesAV,Lin298Dielectric,arbabi2023advances,dorrah2022tunable}. Existing metasurfaces offer flat optics, however, but suffer from severe monochromatic and chromatic aberrations when acting as imagers (\ie, metalens)~\cite{khorasaninejad2017metalenses}. Although dispersion-engineered achromatic metalenses have been reported~\cite{Shrestha2018BroadbandAD,khorasaninejad2017achromatic,Wang2018ABA}, they are fundamentally limited to aperture sizes of tens of microns~\cite{Presutti2020FocusingLimits}. To enable larger aperture size while suppressing aberrations, researchers have designed meta-optic systems together with conventional deconvolution methods~\cite{colburn2018metasurface} and learned recovery methods~\cite{neural_nanooptics} for multi-wavelength color imaging. \rev{Recent metalens camera designs have explored deep neural network reconstruction for analytical hyperbolic metalenses~\cite{dong2024achromatic}, aperture-stop optimization for three discrete wavelengths~\cite{park2025end}, and folded metalens to reduce system thickness~\cite{kim2024metasurface}. While achieving high-quality imaging results, these metasurface cameras are not collaborative designs for broadband imaging, and the number of optimized spectral bands is two orders of magnitude smaller than ours.} The work most related to ours is~\cite{chakravarthula2023thin}, which utilizes an inverse-designed on-sensor metasurface array camera with a learned optimization algorithm for large FOV broadband imaging. While previous work used metalens arrays to expand the field of view at the cost of image quality, our approach jointly designs six complementary metalenses that collaborate for high-fidelity imaging, with a compact \SI{3.6}{\milli\metre} height.

\paragraph{Differentiable Optics.}
While conventional camera design hand-engineers the optics in isolation for compartmentalized metrics (\eg, root-mean-square spot size), a body of work in computational imaging leverages the modern differentiable framework to jointly optimize optics and reconstruction algorithms for downstream tasks.   
Successful methods that follow this paradigm learn optics and/or sensors for color imaging~\cite{Chakrabarti2016LearningSM,sun2021end,wang2022differentiable,yang2024curriculum,yang2024end,na2024end}, microscopy~\cite{Horstmeyer2017ConvolutionalNN,kellman2019data,Nehme2019DenseTD,Shechtman2016MulticolourLM}, depth imaging~\cite{Chang2019DeepOF,Haim2018DepthEF,Wu2019PhaseCam3DL,ghanekar2024passive}, super-resolution and extended depth of field~\cite{Sitzmann2018EndtoendOO,tan2021codedstereo}, time-of-flight imaging~\cite{Marco2017DeepToFOR,Su2018DeepET,chugunov2021mask}, high-dynamic range imaging~\cite{Sun_2020_LearnedOpticHDR,metzler2019deep,martel2020neural,shi2024split}, compressive sensing~\cite{yoshida2018joint,iliadis2020deepbinarymask}, active-stereo imaging~\cite{baek2021polka}, hyperspectral imaging~\cite{baek2021single,fu2020joint,shi2024learned}, and computer vision tasks~\cite{Tseng2021DeepCompoundOptics,souza2024latent,wei2024spatially}. Our approach models differentiable optics for metasurfaces and jointly optimizes the metasurface design and the reconstruction algorithm end-to-end for general broadband imaging.
\begin{figure*}[th!]
\centering
\includegraphics[width=0.95\linewidth]{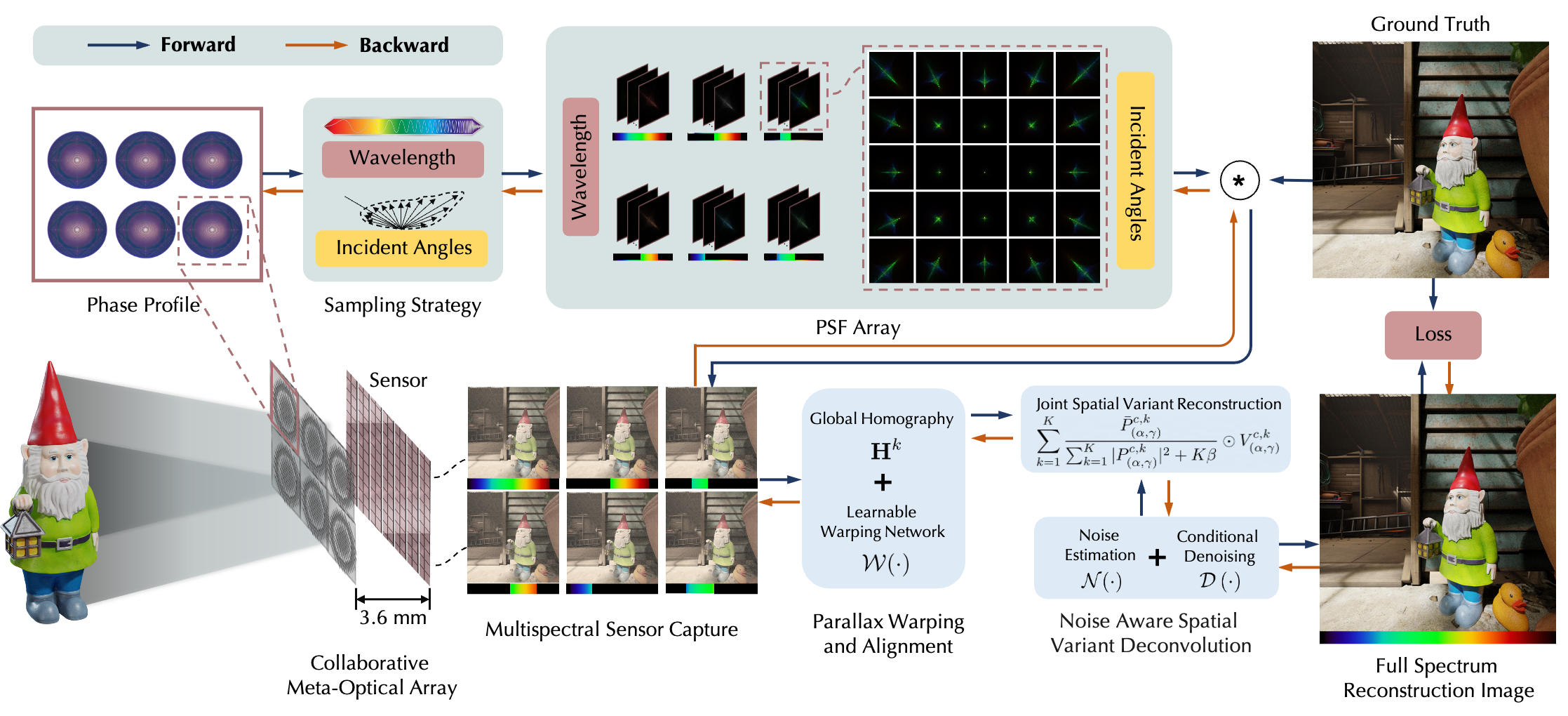}
\vspace{-5mm}
\caption{\textbf{Collaborative Nanophotonic Imaging.} We propose a $2\times 3$ collaborative metalens array camera wherein individual metalenses in the array capture a part of the whole visible spectrum (the spectrum under each sublens image here is for illustration). After 
two-step alignment to compensate for parallax, we reconstruct the full visible spectrum at the resolution
of the input sub-images, with a multi-image variant of Wiener deconvolution collaboratively combining wavelength spectra across the six collaborative measurements. The whole pipeline is fully differentiable, thereby enabling end-to-end optimization of the collaborative nanophotonic imaging system. }
\label{fig:method_overview}
\end{figure*}

\section{Collaborative Nanophotonic Array Imaging}
\label{sec:method}


Our collaborative metalens camera is described in design and reconstruction phase. We first describe the design hinging on a differentiable spatially-varying image formation model with neural metasurface proxy (Section~\ref{sec:formationmodel}), and a joint optimization of metalens array phase profiles and PSFs for multi-image deconvolution (Section~\ref{sec:Imaging}) through distributed metasurface learning (Section~\ref{sec:lmom}). For experimental reconstruction, we devise an array image alignment approach, followed by the same multi-image deconvolution method combined with a learned denoiser (Section~\ref{sec:alignment}). Implementation and training details are provided in Section~\ref{sec:details}. Figure~\ref{fig:method_overview} illustrates our collaborative nanophotonic imaging system design and reconstruction architecture.

In what follows, continuous variables (such as wavelength $\lambda$) are denoted in subscript, while discrete (categorical) variables (such as the metalens index $k$ and color channel $c$) are denoted in superscript.

\subsection{Spatially-varying Image Formation Model}
\label{sec:formationmodel}

\paragraph{Overview of Image Formation Model}
We start by deriving the formation model of an image by a single meta-optical element.
We target broadband imaging in the visible range with wavelength $\lambda$ from \SI{400}{\nano\metre} to \SI{700}{\nano\metre}.
The optical imaging process at a certain wavelength can be generally modeled as the locally-varying convolution of the focal scene image $u_\lambda$ with a point spread function (PSF) $p_\lambda$. 
This PSF kernel is a function of the wavelength-dependent phase profiles $\Phi_\lambda$ in the lens plane\footnote{Note we omit the spatial/angle dependency of the PSF in the formula here for simplicity. The spatial/angle-dependent PSF computation will be detailed later.}
\begin{equation}\label{eq:psf_function_of_phase}
    p_\lambda: \Phi_\lambda \mapsto p_\lambda(\Phi_\lambda),
\end{equation}
which can be derived by a free-space wave propagation of a wavefront (plane wave modulated by the optical phase delay) propagated to the sensor plane~\cite{goodman2005introduction}. 
Our design leverages a $2 \times 3$ array of $K=6$ metalenses, and thus has as many PSFs $p^k_\lambda$ and related phases $\Phi^k_\lambda$. \rev{We chose the number of lenses to align with the sensor aspect ratio; however, the method itself is agnostic to lens number and can accommodate any array configuration.}

The $k$-th image $v^{(k, c)}$ at a certain color channel $c \in \{R, G, B\}$ focused by the optics onto the sensor plane is formed by integrating PSFs over the visible spectrum weighted by the spectral sensitivities of the on-sensor color filters
\begin{align}\label{eq:formationmodel}
    v^{(k, c)} &= \int_{\lambda\in\Lambda}\! \kappa^{c}_{\lambda} \left(u_\lambda \otimes p^k_\lambda\right) \,\mathrm{d}\lambda + \varepsilon^{(k, c)}, 
\end{align}
where $\otimes$ denotes a locally-varying convolution as the PSF kernel varies across the FOV. Here, $\kappa^c_\lambda$ is the camera spectral response function (a.k.a. camera sensitivity~\cite{jiang2013space,solomatov2023spectral}) for each color filter $c$ atop sensor. As such, $\varepsilon^{k,c}$ denotes the random (shot and read) noise in the sensor measurements, which is modeled as a heteroscedastic Gaussian distribution~\cite{Foi2008PracticalPN}.

Due to the nonlinear phase responses of the nano-pillars across different wavelengths, the associated PSFs can vary substantially among spectral bands~\cite{chen2018broadband}. Consequently, Eq.~\eqref{eq:formationmodel} cannot be approximated by a simple convolution in the RGB image space (see \supp{} for details). To accurately capture the underlying physics, we therefore optimize our optics model in the hyperspectral domain.


Given the smooth variation of the PSFs across different incident angles~\cite{lohmann1965space}, the locally-varying convolution $\otimes$ can be approximated as spatially-uniform convolutions $\ast$ of $M \times N$ image patches localized in $u_\lambda$, with the corresponding PSFs $p_\lambda^{(k, m, n)}$ computed at specific incident angles~\cite{Hirsch2010EfficientFF, schuler2011non, Eboli2022fast}.
The image $u_\lambda\otimes p_\lambda^k$ is then formed by blending these patches with a fusion windowing function $w$ (\eg, a Hann window~\cite{blackman1958measurement}) to suppress border artifacts. This can be formulated as
\begin{equation}\label{eq:localconvolution}
    \left(u_\lambda \otimes p^k_\lambda\right) = \sum_{m=1}^M\sum_{n=1}^N w^{(m,n)} \left(u_\lambda * p_{\lambda}^{(k, m, n)}\right),
\end{equation}
where $(m,n)$ indicates the $(m,n)$-th entry of an array of varying incident angles. The window $w^{(m, n)}$ is non-zero only for the interval of the corresponding patch $(m,n)$ in $u_\lambda$.
In what follows, we elucidate Eq.~\eqref{eq:psf_function_of_phase} in detail, which includes the metalens model and the angle-dependent PSF computation. 

\paragraph{Metalens Model Parameterization.}
Each individual metalens is composed of silicon nitride nanoposts with a height of $\SI{1000}{nm}$ atop a fused silica substrate.
These nanostructures with optimizable width $D$ between $100$-$\SI{300}{nm}$ are distributed on a regular grid with $P = \SI{350}{nm}$ distance, resulting in a grid of pillars with tunable width (which determines the ``duty cycle'' $d = \frac{D}{P}$ between 0 to 1)~\cite{neural_nanooptics, chakravarthula2023thin}.
Instead of working on this low-level representation that entails the computationally demanding full-wave solver, 
 we parameterize each metalens as a radially symmetric phase profile 
\begin{equation}
    \Phi_{\lambda_0} (x, y) = \varphi\left( r \right),
    \quad
    r = \sqrt{x^2 + y^2},
\label{eq:radially-symmetric-phase}
\end{equation}
where $(x, y)$ represents the spatial coordinates in the lens plane, and $\varphi^k$ is a 1D radial vector of size $L$ (\ie, the radius of the metalens), which serves as the optimization variable in our design.
Note the above phase function is only defined for a specific nominal wavelength $\lambda_0$. We select $\lambda_0=\SI{658}{nm}$ as the nominal wavelength for our fabricated structure\rev{, as we empirically found that, there exists a one-to-one (instead of one-to-many) mapping relationship between the structure and the phase response, within a tolerance for the wavelength. }

To fully characterize the broadband behavior of the metalens while maintaining the differentiability of the pipeline, we utilize two proxy functions in sequence to determine the phase profiles of the metalens across the entire visible spectrum of interest in lieu of the rigorous yet computationally expensive full-wave methods, such as the finite-difference-time domain method (FDTD)~\cite{kunz1993finite, taflove2005computational}. 

The first proxy is a precomputed polynomial that computes the metasurface geometry (duty cycle $d$ for each nanopost) from the phase at nominal wavelength $\lambda_0$ as $d(\phi_{\lambda_0}) = \sum_{o=0}^O a_o \left(\phi_{\lambda_0} / (2\pi)\right)^{o}$, where we find that a fifth-order polynomial $O = 5$ is sufficient to fit the phase-to-structure mapping data in experiments.

\begin{figure}[t]
    \centering
    \begin{overpic}[width=1.0\columnwidth]{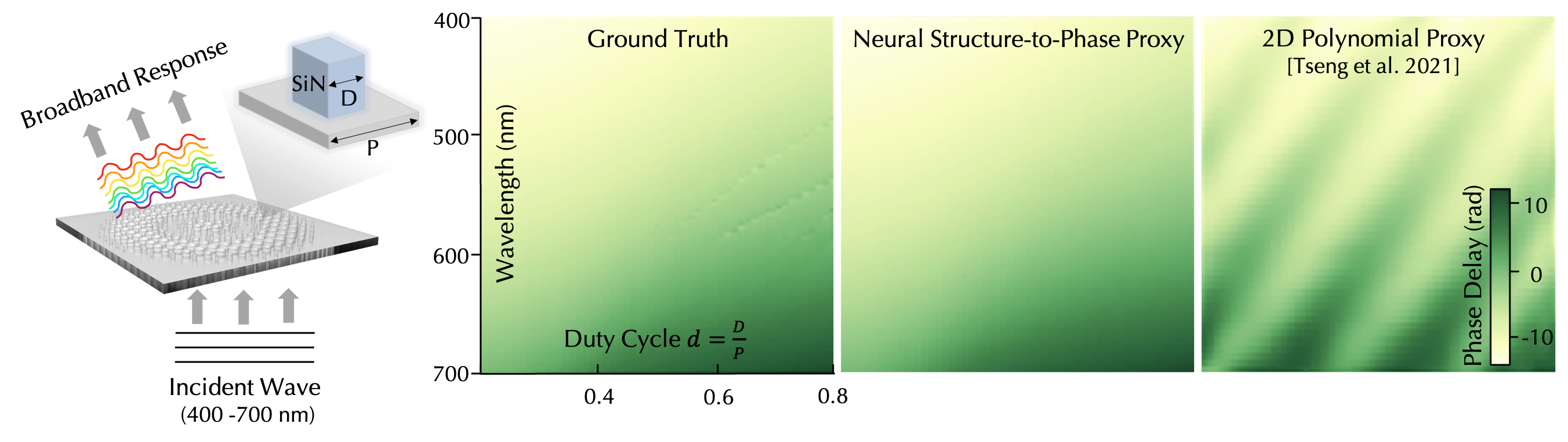}
    \end{overpic}
    \vspace{-12pt}
    \caption{\textbf{Comparison between proposed neural structure-to-phase proxy and polynomial proxy~\shortcite{neural_nanooptics}.} A proxy is needed to model the broadband response of a metalens given plane wave illumination across visible wavelengths. For a single nano-pillar shown on the left, its induced phase delay is determined by both the incident wavelength (vertical axis) and the duty cycle of the structure (horizontal axis). We plot the ground truth of the unwrapped phase delay as well as the approximated values from two differentiable proxy models. In contrast to the existing polynomial proxies~\shortcite{neural_nanooptics}, the proposed neural proxy accurately models the nonlinear relationship among nano-structure duty cycle, target wavelength, and broadband phase delay.}
    \label{fig:proxy_comparison}
\end{figure}

Due to the complex, nonlinear relationship among the nano-structure duty cycle, target wavelength, and the broadband phase delay—together with the high-precision requirements of broadband phase mapping—polynomial based proxy functions used in previous literature~\cite{neural_nanooptics, chakravarthula2023thin} fail to converge within a \SI{1}{\percent} error threshold, see Fig.~\ref{fig:proxy_comparison}. Therefore, we utilize a neural network proxy ${\mathcal{N}_\textit{proxy}}$ as the second structure-to-phase mapping proxy to compute the phases at other target wavelengths $\phi_t$ given the metasurface duty cycle $d$. 
\begin{equation}
    \Phi_{\lambda_t}(d) = \mathcal{N}_{\textit{proxy}}(d, \lambda_t) \bmod 2\pi.
\label{eq:duty-to-phase}
\end{equation}
The neural network comprises three fully connected layers with ReLU activation. Its training dataset is generated by FDTD simulations at \SI{2}{\nano\metre} increments for structure sizes ranging from \SI{80}{\nano\metre} to \SI{280}{\nano\metre}, and at \SI{1}{\nano\metre} increments across the visible spectrum. With the two proxy functions introduced, we now build a differentiable computational graph $ \Phi_{\lambda_0} \mapsto \Phi_{\lambda_t}$ from the phase profile at nominal wavelength $\lambda_0$ to other target visible wavelengths $\lambda_t$. See the \supp{} for full details of the FDTD simulation.

\paragraph{Angular-dependent PSF Model.}
With the metalens model, we can now derive the angle-dependent PSFs of the whole optical system. 
Given an incident wave direction (altitude $\alpha$ and azimuth $\gamma$ angles), the incident plane wave phase  $\phi^{(\alpha,\gamma)}_{\lambda}$ at wavelength $\lambda$ is given by
\begin{equation}
    \phi^{(\alpha,\gamma)}_{\lambda}(x, y) = \frac{2\pi}{\lambda}\frac{x \tan\alpha + y \tan\gamma}{\sqrt{\tan^2 \alpha + \tan^2 \gamma + 1}}.
    \label{eq:sv-input}
\end{equation}
As explained before, the PSF smoothly varies across the FOV and thus we split the whole FOV into $M \times N$ patches, each of which corresponds to a certain angular position ($\alpha[m, n]$, $\gamma[m, n]$).
In practice, we discretize the FOV into a $7\times 7$ array, with angles spanning equally from $-\ang{30}$ to $\ang{30}$. 
As such, the modulated phase function $\Phi^{(k,m,n)}_\lambda$ (with a little abuse of notations) in the lens plane is
\begin{equation}
    \Phi^{(k,m,n)}_\lambda(x, y) = \Phi_\lambda^k(x, y) + \phi_\lambda^{(\alpha[m,n], \gamma[m,n])} (x, y).
\end{equation}
Finally, the destination field (also its intensity, \ie, PSFs) in the sensor plane can be computed by Rayleigh-Sommerfeld scalar diffraction integral~\cite{goodman2005introduction} of the source field $e^{j\Phi}$ in the lens plane, numerically implemented by a (band-limited) shifted angular spectrum method (ASM)~\cite{matsushima2010shifted}\footnote{Note the shifted ASM kernel also depends on the incident angles. The details are deferred to the \supp{}.}.

Our image formation model is fully differentiable since all operations in Eq.~\eqref{eq:formationmodel} are differentiable. We implement this in an automatic-differentiation framework, enabling backpropagation~\cite{lecun2015deep} and gradient-based optimization for the inverse design of the nanophotonic array camera.

\subsection{Learning Collaborative Metalens Arrays}
\label{sec:Imaging}
\paragraph{Broadband Achromatic Initialization}
With the image formation model described in Section \ref{sec:formationmodel}, we propose a broadband metalens phase initialization for the later joint optimization process. To this end, we optimize a rotationally symmetric metalens phase to maximize the center energy of the PSFs under all sampled incident wavelengths $\Lambda_\text{disc} $ and angles $ \alpha_\text{disc}, \gamma_\text{disc}$. Formally, we solve the following optimization
\begin{equation}
\label{eq:psf-ini}
    \widetilde{\phi}_{\lambda_0}(r) = \argmax_{\phi_{\lambda_0}}
    \sum_{\lambda_t \in \Lambda_\text{disc}} \sum_{\alpha \in \alpha_\text{disc}} \sum_{\gamma \in \gamma_\text{disc}}
    p_{\lambda_t}^{(\alpha, \gamma)}(\Delta x_{p^{(\alpha, \gamma)}}, \Delta y_{p^{(\alpha, \gamma)}}),
\end{equation}
where
\begin{equation*}
\begin{aligned}
     \Lambda_\text{disc} & = \{ \lambda \mid \lambda = 400 + 0.5k, \, k \in \mathbb{Z}, \, 400 \leq \lambda \leq 700 \},\\
     \alpha_\text{disc} & = \{ \alpha \mid \alpha = 0 + 5k, \, k \in \mathbb{Z}, \, 0 \leq \alpha \leq 30 \},\\
    \gamma_\text{disc} & = \{ \gamma \mid \gamma = 0 + 5k, \, k \in \mathbb{Z}, \, 0 \leq \gamma \leq 30 \},
\end{aligned}
\end{equation*}
and $(\Delta x_{p^{(\alpha, \gamma)}}, \Delta y_{p^{(\alpha, \gamma)}})$ denote the offsets of the PSFs center under incident wave direction $(\alpha, \gamma)$ on axis $X$ and axis $Y$. By geometry, the offsets are proportional to focal length $f$ and the incident angles
\begin{equation}
\label{eq:psf-center-shift-ang}
\bigl( (\Delta x_{p^{(\alpha, \gamma)}}, \Delta y_{p^{(\alpha, \gamma)}})
\;=\;
\bigl(-f\tan\alpha,\, -f\tan\gamma \bigr).
\end{equation}
This starting phase ensures high initial focusing performance across the entire visible spectrum (sampled every \SI{0.5}{nm}) and for both on-axis and off-axis incident angles ranging from \SI{0}{\degree} to \SI{30}{\degree} (sampled every \SI{5}{\degree}). 

\paragraph{Joint Optimization with Multi-image Deconvolution}

Following the broadband phase initialization, we jointly optimize six metalenses in tandem with a multi-image joint deconvolution algorithm. As in Eq.~\eqref{eq:psf-ini}, we employ the stochastic sampling technique from Section~\ref{sec:lmom} to discretize the continuous visible spectrum \(\Lambda\) into the set \(\Lambda_{\text{disc}}\). This effectively approximates the continuous image formation model for the color channel $c$ in Eq.~\eqref{eq:formationmodel} via
\begin{align}\label{eq:discrete_image_formation} 
    v_{(\alpha, \gamma)}^{k, c} = \sum_{\lambda \in \Lambda_\text{disc}} \kappa^{c}_{\lambda} \left(u_\lambda \otimes p^k_{\lambda, (\alpha,\gamma)}\right) + \varepsilon^{(k, c)}, 
\end{align}
where the hyperspectral latent images \(u_{\lambda}\) are convolved with the corresponding hyperspectral PSFs \(p_{\lambda, (\alpha, \gamma)}^{k}\), under the incident direction \((\alpha, \gamma)\), plus camera noise \(\varepsilon^{(k,c)}\). 

During lens array training, we initially disregard potential parallax effects stemming from the array design to focus on improving the optical properties. Consequently, the reconstruction problem reduces to a multi-measurement, single-object scenario. In order to design an ultra-compact camera with minimal computational overhead, we adopt a multi-frame deconvolution approach~\cite{yaroslavsky1994deconvolution}, which jointly deconvolves the \(K\) frames and returns a single sharp result by solving for a given wavelength $\lambda$ 
\begin{equation}\label{eq:multiframeLS}
    \min_{u_\lambda}\frac{1}{K}\sum_{k=1}^K  
    \| v_{\lambda, (\alpha, \gamma)}^{k} - u_\lambda \otimes p^k_{\lambda, (\alpha,\gamma)} \|_2^2 + \beta \|u_\lambda\|^2_2,
\end{equation}
with the following closed-form solution in the Fourier domain
\rev{\begin{equation}\label{eq:deconvfilter}
    U_{\lambda} = 
    \sum_{k=1}^K\frac{\overline{P}^{k}_{\lambda, (\alpha, \gamma)}}{{\sum_{\ell=1}^K | P^{\ell}_{\lambda, (\alpha, \gamma)}|^2 + K \beta}} \odot V^{k}_{\lambda, (\alpha, \gamma)},
\end{equation}}
where uppercase letters denote the Fourier transforms of the corresponding lowercase variables, and \(\overline{P}\) is the complex conjugate of \(P\). Multiplication (\(\odot\)) and division are applied element-wise. We set the regularization weight $\beta$ to 0.001.

Equation~\eqref{eq:deconvfilter} averages the $K$
measurements processed with individual filters akin to the original Wiener filter~\cite{Wiener1964} but
where the denominator features the sum of the moduli of the $K$ PSFs
instead of a single one as in the original formulation.
The individual filters are consequently aware of the other measurements during joint deconvolution into a single sharp result.
We call {\em joint Wiener filter} this global filter applied to the $K$ images $v^{k}_{\lambda, (\alpha, \gamma)}$.
See the \supp{} for detailed derivation and explanations.

The joint Wiener filter enforces collaborative behavior across the lens array at each target wavelength \(\lambda_t\). Rather than constraining each individual lens to perform uniformly well over the entire spectrum, as in Eq.~\eqref{eq:psf-ini}, the contribution of each lens in the Fourier domain is weighted by its individual signal-to-noise ratio (SNR). 
This design relaxes the broadband capabilities of each lens and promotes mutual compensation, resulting in an improved broadband reconstruction when all lenses work collaboratively as a group.

However, the hyperspectral measurements $V^{k}_{\lambda, (\alpha, \gamma)}$ required in Eq.~\eqref{eq:deconvfilter} are challenging to acquire in commodity RGB color filter array (CFA) sensors. As such, we instead approximate the previous equation with the following color-channel broadband deconvolution method
\rev{\begin{equation}
    \label{eq:rgb_joint}
    U^c_{RGB} = \sum_{k=1}^K\frac{\widebar{P}^{c, k}_{(\alpha, \gamma)}}{{\sum_{\ell=1}^K| P^{c, \ell}_{(\alpha, \gamma)}|^2 + K \beta}} \odot V^{c, k}_{(\alpha, \gamma)}.
\end{equation}}
We employ here instead the camera spectral response function (CSRF) weighted broadband average PSF and RGB image as
\begin{equation}
    \label{eq:rgb_psf_measurement}
    P^{c,k}_{(\alpha,\gamma)} = \sum_{\lambda \in \Lambda_\mathrm{disc}}\kappa^{c}_{\lambda}P^k_{(\alpha,\gamma)} \quad \text{and} \quad
    V^{c,k}_{(\alpha,\gamma)} = \sum_{\lambda \in \Lambda_\mathrm{disc}}\kappa^{c}_{\lambda}V^k_{(\alpha,\gamma)}.
\end{equation}
The CSRF weights $\kappa^{c}_{\lambda}$ from Eqs.~\eqref{eq:formationmodel} and \eqref{eq:discrete_image_formation} combine all the wavelengths $\lambda$ in $\Lambda_\mathrm{disc}$ within a single joint Wiener filter.

\begin{figure*}[t]
    \vspace*{-6pt}
    \centering
    \includegraphics[width=0.91\linewidth]{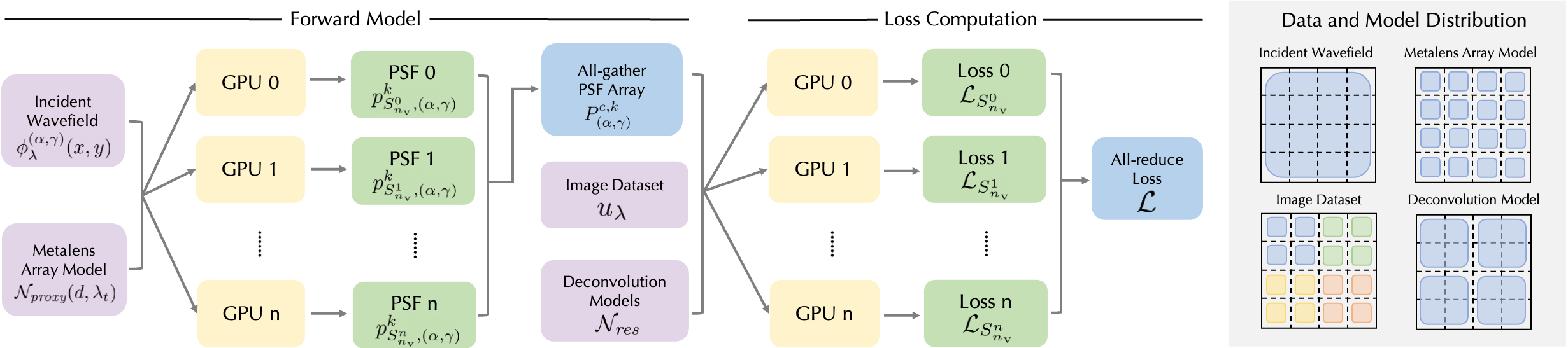}\vspace*{-6pt}
    \caption{\textbf{Distributed large meta-optics model optimization pipeline with Full Sharded Data Parallel (FSDP) illustration.} In the forward model, the incident wavefield is distributed across multiple GPUs. Each GPU has a complete metalens array model and computes the PSF for its assigned wavelengths, then contributes to the global PSF array through an all-gather operation. When computing loss, the image dataset is also distributed across GPUs. Depending on the size of the deconvolution model, we can also distribute the model weights. Each GPU calculates its local loss, and these losses are combined through an all-reduce operation to compute the global loss for optimization. The right figure provides a detailed illustration of data and model distribution, where the blocks represent a $4 \times 4$ GPU cluster and each color block represents either a complete dataset or model weights. The different colors in the left bottom figure represent different image datasets. 
    }
    \label{fig:distibuted-pipeline}
\end{figure*}

\paragraph{Spectrally Agnostic Training}
Instead of trying to estimate the spectrum from the measurements, we aim to learn a spectrally agnostic optical design with consistently good broadband performance. 
A way to simulate challenging spectral conditions is to sample a subset of bands during training randomly.
Specifically, we randomly select \(n_\text{v}\) wavelengths from \(\Lambda_{\text{disc}}\) (denoted as \(S_{n_\text{v}}\)) for the image formation process \(V^{c,k}_{(\alpha,\gamma)}\), and \(n_\text{p}\) wavelengths (denoted as \(S_{n_\text{p}}\)) for constructing the simulated deconvolution PSFs \(P^{c,k}_{(\alpha,\gamma)}\), where \(|S_{n_\text{p}}| \gg |S_{n_\text{v}}|\). In our experiments, we set \(|S_{n_\text{p}}| = 108\) and \(|S_{n_\text{v}}| = 3\). We then construct a simulated measurement



\begin{equation}
    W^{c,k}_{(\alpha, \gamma)} = \sum_{\lambda_t \in P_v}\kappa^{c}_{\lambda}V^{k}_{\lambda, (\alpha, \gamma)},
\end{equation}
which we recover with the approximate joint Wiener filter of Eq.~\eqref{eq:rgb_joint} applied to $W^{c,k}_{(\alpha, \gamma)}$ in place of $V^{c,k}_{(\alpha, \gamma)}$ as synthetic input measurement.
The loss function for optical training reads:
\begin{equation}
\label{eq:lens-loss}
    \mathcal{L}_\mathrm{optics} = \| U^c_{RGB} -\displaystyle\sum_{\lambda_t\in S_{v}}\kappa_{\lambda_t}^{c}U_{\lambda_t} \|_2^2.
\end{equation}
Equation~\eqref{eq:lens-loss} implicitly aligns the narrow-band RGB PSFs with their broadband counterparts while promoting the broadband collaborative PSFs that provide high reconstruction quality. This training will also make the lens robust to various illumination sources in the real world, owing to the random sampling policy.


\rev{Optimizing PSFs for a 100-million nano-post metasurface at 108 wavelengths demands significant GPU memory (around 2.1 TB intermediate tensors)}. To address this challenge, we leverage the distributed training framework detailed in Section~\ref{sec:lmom}. Each local device independently samples \(S_{n_\text{v}} \sim \Lambda_{\text{disc}}\) and then utilizes an \texttt{all-gather} operation to obtain the narrow-band PSFs from other devices, thereby forming the global broadband PSF 
\(P^{c,k}_{(\alpha,\gamma)}\). Section~\ref{sec:lmom} provides additional details on our distributed implementation.

\subsection{Image Reconstruction Pipeline}
\label{sec:alignment}


\paragraph{Alignment under Parallax.}
Our metalens array captures $K$ subimages \(v^{k}\) ($k=1,\dots,K$) of the same scene 
from slightly different viewing 
angles, thereby inducing parallax between the subimages. We propose a 
two-stage strategy to align these subimages to a common reference 
image~\(\widetilde{v}\), which we choose to be the top middle subimage.
We first coarsely align the \(v^{k}\)'s to a common
system of coordinates with \textit{global homographies}. We obtain
the $3\times3$ homography matrices \(\mathbf{H}^{k}\) by calibrating 
the $K$ apertures with a ChArUco broad pattern. Each 
subimage \(v^{k}\) is 
warped to the reference system as
\begin{equation}
    v_{\mathrm{coarse}}^{k}(q)
    \;=\;
    v^{k}\bigl(\mathbf{H}^{k}\,q\bigr),
\end{equation}
where \(q\) denotes pixel coordinates.
Local misalignments may remain between the \(v_{\mathrm{coarse}}^{k}\)
and the reference due to depth variations or 
lens-specific distortions that a single homography cannot capture.
Hence, we refine the previous coarse displacement with dense 
optical flows \(\mathbf{D}^{k}(u)\in\mathbb{R}^2\) with a pretrained RAFT model~\cite{teed20raft}, denoted by \(\mathcal{W}\).
Since the coarsely-aligned images are aberrated, 
\(\mathcal{W}\) trained on sharp natural images predicts inaccurate flow maps. 
We thus opt to first deconvolving
the images \(v_{\mathrm{coarse}}^{k}\) with individual Wiener filters reusing the 
Fourier transforms of the broadband PSFs in Eq.~\eqref{eq:rgb_psf_measurement}. Let
\(s_{\mathrm{coarse}}^{k}\) (resp.~\(\widetilde{s}\)) be the deconvolved version of \(v_{\mathrm{coarse}}^{k}\) (resp.~\(\widetilde{v}\)), the estimation of the flow maps
\(\mathbf{D}^{k}\) thus reads
\begin{equation}
    \mathbf{D}^{k}
    \;=\;
    \mathcal{W}\bigl(
        s_{\mathrm{coarse}}^{k},\,
        \widetilde{s}
    \bigr).
\end{equation}
Lastly, we warp the blurry subimages to the reference following
\begin{equation}
    v_{\mathrm{warp}}^{k}(q) = v_{\mathrm{coarse}}^{k}
    \bigl(
        \mathbf{D}^{k}(q)
    \bigr).
\end{equation}

\paragraph{Joint Wiener Filtering.}
After having aligned the frames, we recover a single sharp image
with a joint Wiener filter.
To account for the spatially-varying formation model in Eq.~\eqref{eq:discrete_image_formation}, we first fragment the images $v_{\mathrm{warp}}^{k}$ into
overlapping patches centered at locations $(m,n)$. We apply Eq.~\eqref{eq:rgb_joint} to combine the $K$
patches into a single one at each patch location $(m,n)$ in the Fourier domain.
Lastly, we merge the patches into a single restored sharp image $\widehat{u}_\mathrm{fused}$. To prevent edge artifacts during patch merge, we apply a Hann window $w^{(m,n)}$ to the deconvolved patches before the fusion.

\paragraph{Noise-Aware Image Denoising.}
The joint Wiener module is a high-pass filter that removes blur, but
that also boosts the noise from the measurements resulting in
correlated noise in $\widehat{u}_\mathrm{fused}$. Furthermore, because of the wavelength-dependent PSFs
of metalens, the boosted noise level also varies with the scene spectrum distribution.  

Consequently,
we develop a \emph{noise-aware} post-fusion reconstruction module that integrates noise 
estimation and conditional denoising to further enhance image quality.

We employ a residual denoising neural network \(\mathcal{F}\) that 
denoises the fused image $\widehat{u}_\mathrm{fused}$. First, a 
\emph{noise-estimation} subnetwork \(\mathcal{N}\) predicts an estimation
of the 
variance map \(\widehat{\sigma}\) as:
\begin{equation}
    \widehat{\sigma}
    \;=\;
    \mathcal{N}\bigl(\widehat{u}_\mathrm{fused}\bigr).
\end{equation}
Subsequently, this noise estimate serves as guidance to a \emph{conditional denoising} subnetwork \(\mathcal{D}\) as:
\begin{equation}
    \widehat{u}
    \;=\;
    \mathcal{D}\Bigl(
        \widehat{u}_\mathrm{fused},
        \widehat{\sigma}
    \Bigr),
\end{equation}
where \(\widehat{u}\) represents the final reconstructed and denoised 
estimate of the scene. By providing explicit noise information, $\mathcal{D}$
becomes a non-blind denoiser that is adaptive to the input image, 
thus preserving fine structures in low-noise regions and 
suppressing artifacts in high-noise areas.

We first derive a closed-form solution to estimate the noise variance after joint Wiener deconvolution, leveraging the hyperspectral PSFs of the lens array to train our noise estimator subnetwork \(\mathcal{N}\).  
(See \supp{} for details of the derivation.)  
Subsequently, we train our noise-aware reconstruction model \(\mathcal{D}\) using a combination of supervision and regularization terms:
\begin{equation}
    \mathcal{L}
    \;=\;
    \mathcal{L}_{\mathrm{MSE}}
    \;+\;
    \alpha\,\mathcal{L}_{\mathrm{\mathrm{TV}}},
\end{equation}
where $\mathcal{L}_{\mathrm{MSE}}$
is the mean-square error between the prediction
of the network and a ground-truth image, and
$\mathcal{L}_{\mathrm{\mathrm{TV}}}$ denotes the total variation \cite{rudin1992nonlinear} loss, \(\alpha\) is a weighting parameter. Specifically, we adopt $ \alpha = 0.001$ to trade off fidelity and local smoothness.


\subsection{Distributed Large Meta-Optics Optimization}
\label{sec:lmom}
Training a meta-optics array with 100 million parameters under one-hundred distinct wavefield conditions can easily require up to 2 TB of GPU memory—an order of magnitude exceeding the capacity of a single GPU. As a result, existing broadband meta-optics designs~\cite{neural_nanooptics, chakravarthula2023thin, froch2025beating} have been constrained to smaller aperture sizes, fewer spectral bands, and on-axis PSFs, largely due to limited GPU memory.

To relieve this constraint, we develop a distributed large meta-optics model optimization pipeline with Full Sharded Data Parallel (FSDP)~\cite{zhao2023pytorch}, as illustrated in Fig.~\ref{fig:distibuted-pipeline}.  We distribute the incident wavefields, specified by the wavelength $\lambda $ and incident angles $(\alpha, \gamma)$ across multiple GPUs, akin to the concept of data in the machine learning context. We adopt a stochastic sampling scheme analogous to mini-batch training in deep learning. Rather than fixing a discrete set of wavelengths as in prior work \cite{chakravarthula2023thin}, each optimization iteration draws a subset of wavefields randomly sampled across various $\lambda$ and $(\alpha, \gamma)$ values. This randomization ensures broad coverage of the underlying high-dimensional design space over the course of training, providing more robust optimization outcomes and less susceptibility to overfitting any particular set of wavefields.

Although a single GPU can generally manage the total number of trainable parameters for a single meta-optics model, the intermediate tensors generated during PSF computations for each wavefield are often two orders of magnitude larger than the model parameters themselves.
To mitigate this overhead, we shard the model across multiple GPU ranks and split the wavefield inputs accordingly. Leveraging FSDP in PyTorch, we effectively partition both the model and wavefield data, ensuring that each GPU processes only a fraction of the intermediate tensors.

While FSDP manages parameter and gradient synchronization across GPUs, large-scale broadband meta-optics training introduces additional global synchronization demands. 
For example, broadband PSF calculation during training with Eqs.~\eqref{eq:rgb_joint} and~\eqref{eq:lens-loss} necessitates an \emph{all-gather} operation to share each rank’s PSF results with all other ranks. Similarly, when computing a broadband loss term (\eg, Eq.~\eqref{eq:psf-ini}), an \emph{all-reduce} operation is required to accumulate partial losses from individual ranks into a global loss value. Our proposed training pipeline incorporates a suite of meta-optics-specific utilities—built on top of PyTorch distributed collectives—to address these synchronization requirements (details in \supp{}). These utilities cleanly integrate with distributed optimizers, ensuring that gradient information remains consistent with model parameters throughout the entire training process. The result is a fully end-to-end differentiable framework for large-scale meta-optics design. 

\subsection{Implementation and Training Details}
\label{sec:details}

\paragraph{Meta-Optics Optimization}
We utilize the proposed distributed large-scale meta-optics framework in conjunction with a hyperspectral dataset \cite{jeon2024spectral} to conduct optics optimization. Specifically, we deploy the framework by distributing metalens array model instances across 36 Nvidia H100 GPUs (each with 80\,GB of memory) on 9 nodes (4 GPUs per node), using a data-parallel training scheme. Each rank (i.e., each GPU) receives distinct incident wavefields---characterized by unique wavelengths and incident angles---sampled via the distributed sampler. Due to the considerable volume of intermediate results generated during wave propagation, each rank can only process three incident fields at a time. Consequently, a total of 108 field conditions (36 ranks $\times$ 3 incident fields per rank) are evaluated during each training iteration.


By leveraging the hyperspectral dataset, each rank performs a shift-invariant convolution weighted by the hyperspectral illumination to generate the hyperspectral array measurement, which is then converted into RGB space using the camera’s spectral response function. Noise is subsequently introduced, and the resulting hyperspectral array measurements are passed to the joint Wiener deconvolution module to compute the loss. This loss signal is then backpropagated to update the metalens parameters, while an all-reduce operation ensures that the optimized parameters remain consistent across all ranks.

\paragraph{Reconstruction Pipeline Training}

Given a jointly optimized array design, we train the reconstruction pipeline in a sequential training stage. Specifically, we aim to learn an inverse reconstruction model that jointly tackles parallax alignment, large off-axis astigmatism correction, and denoising on our proposed hyperspectral parallax dataset introduced in Section~\ref{subsec:hs_parallax_dataset}.

We begin by dividing the target scene into $9 \times 9$ patches, corresponding to the sensor dimensions and the spatial arrangement of the metalens array. For each sub-aperture lens, we infer a full-spectrum (400\,nm to 700\,nm in 2\,nm increments) PSF tensor based on the incident angle at the center of each patch. We then convolve each hyperspectral patch with the appropriate spectral PSF tensor, apply window-based blending between adjacent patches, and subsequently convert the resulting measurements back to sensor RGB space with CSRF. This process yields a realistic sensor measurement that accurately simulates parallax and angular aberrations across the entire FOV.


After alignment, see Section~\ref{sec:alignment}, we apply the patch-wise joint Wiener deconvolution followed by residual denoising (initialized with a pretrained image denoiser~\cite{guo2019toward}) to warped sub-aperture images, jointly reconstructing the scene in an end-to-end manner. The overall training loss is backpropagated through the denoising, deconvolution modules, and parallax alignment, allowing the entire pipeline to be optimized for reconstruction fidelity. 

\section{Datasets and Experimental Prototype}
\label{sec:experiment}

%

\begin{figure}[t]
    \vspace*{-6pt}
    \centering
    \includegraphics[width=1.0\columnwidth]{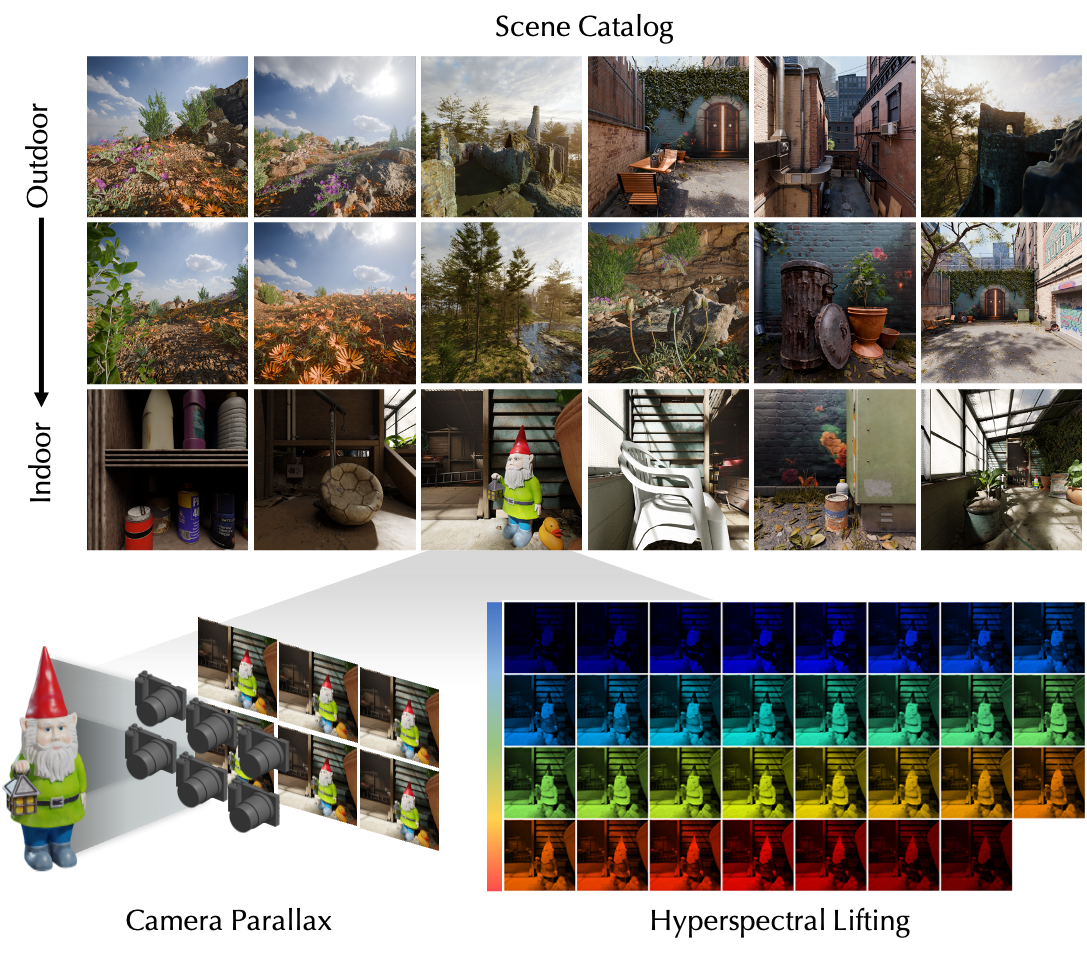}\vspace*{-6pt}
    \caption{\textbf{Illustration of synthetic dataset generation.} The scene catalog shows representative examples of our dataset. For each scene in the dataset, we set up a $2 \times 3$ camera array to capture scene parallax and apply hyperspectral lifting to get images under wavelengths from \SI{400}{\nano\metre} to \SI{700}{\nano\metre}. }    
    \label{fig:dataset}
\end{figure}

\subsection{Synthetic Hyperspectral Parallax Dataset}
\label{subsec:hs_parallax_dataset}
To circumvent the difficulty of capturing large amount of meta-optic array sensor measurements with corresponding high-quality ground truth, we develop a physics-based realistic rendering pipeline that can accurately synthesize realistic metalens array measurements that resemble the real-world observations. Our rendering pipeline here is inspired by today's photorealistic spectral rendering technology developed in the computer graphics community~\cite{jakob2019low,mallett_spectral_2019,van_de_ruit_metameric_2023,otsu_reproducing_2018,meng_physically_2015}, yet we inject an additional rendering layer that imposes spatial-spectral-dependent PSFs to generate highly color-aberrated sensor measurements from otherwise clean images (see Fig.~\ref{fig:method_overview}).  
Specifically, we use the 3-D scenes from Poly Haven~\cite{polyhaven}, a curated public 3-D asset library that contains complete Blender~\cite{blender} 3-D render-ready scenes. 
We set up camera positions to capture objects with different sizes, textures, colors, and distances from the camera.  The sensor size and focal length of the camera are set to match the design of one single metalens camera. We offset the position of the camera according to the distance between metalenses in the array to capture the parallax in different scenes. 
A total of 480 sets of photorealistic linear sRGB images in an array are thus rendered using the Cycles engine~\cite{cycles} under the CIE standard illuminant D65, which covers diverse indoor and outdoor scenes in cities and nature; see Fig.~\ref{fig:dataset} for some representative scenes in our dataset.
These linear sRGB images are then fed into a spectral uplifting method \cite{jakob2019low} to synthesize hyperspectral reflectance data. Next, a spectral illuminant, randomly sampled from a spectral illumination dataset~\cite{li2021multispectral}, is blended with the spectral reflectance to synthesize realistic hyperspectral irradiance images.
Finally, these hyperspectral irradiance images can be optionally convolved with the wavelength- and angle-dependent PSFs (computed from our meta-optic model)\footnote{PSF-based aberration rendering is disabled for the ground truth.} before the spectral integration weighted by a sensor spectral response function~\cite{jiang2013space}, specifically for the Sony IMX 174 sensor in our prototype, to synthesize realistic color sensor observations.

\subsection{Experimental Setup and Datasets Capture}\label{subsec:prototype}
\begin{figure}[t!]
    \centering
    \includegraphics[height=0.45\columnwidth]{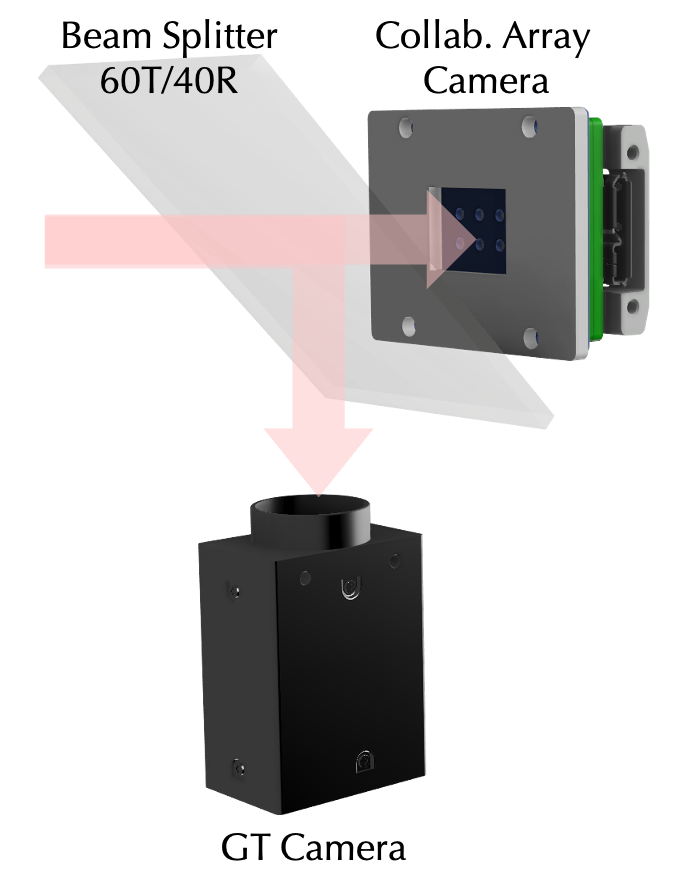}
    \includegraphics[height=0.4\columnwidth]{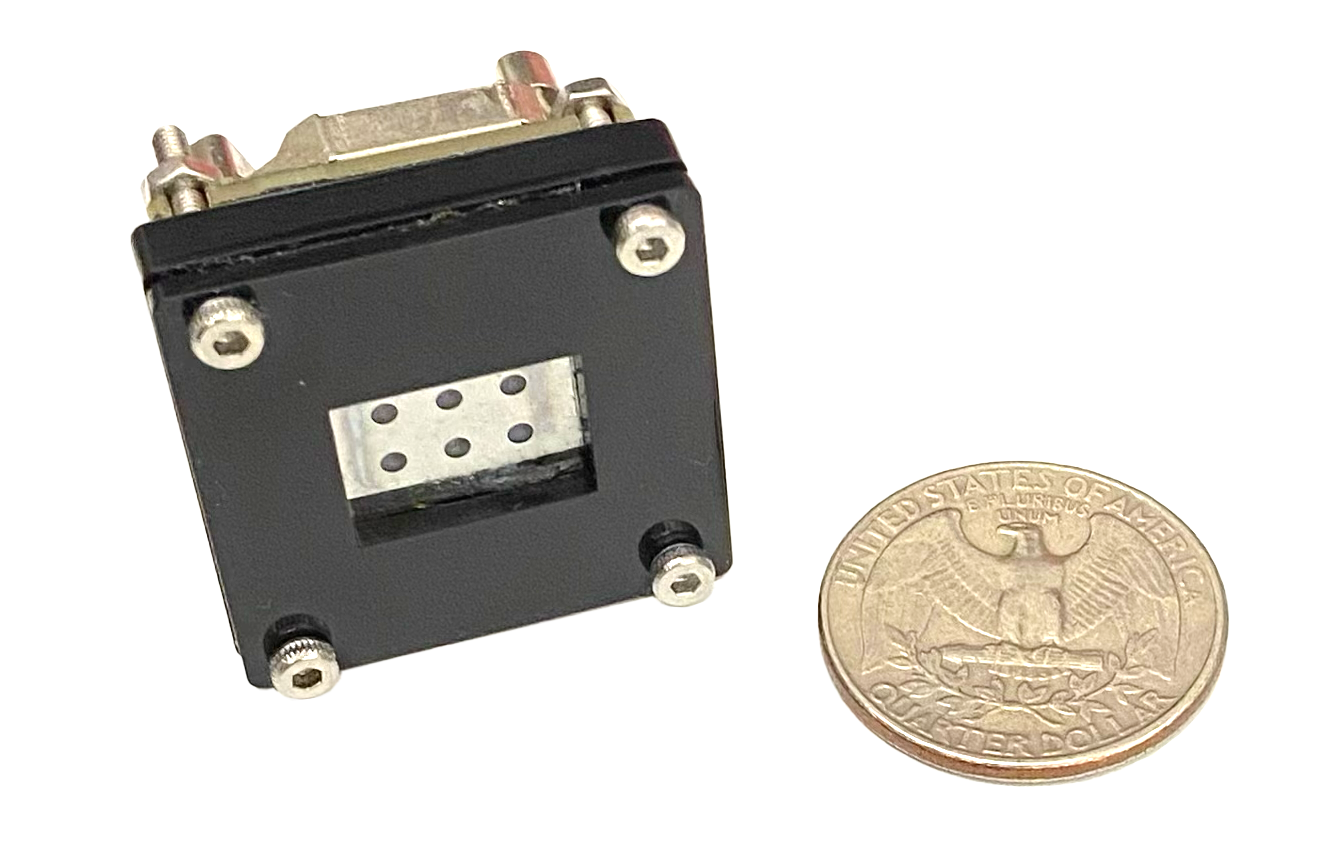}
    \vspace{-3mm}
    \caption{\textbf{Data acquisition and experimental comparisons.} Our setup uses a 40R/60T plate beamsplitter to capture scenes with two cameras at once: a metalens array camera that receives \SI{60}{\percent} of the light passing through the beamsplitter, and a conventional camera (reference camera) that receives \SI{40}{\percent} of the light reflected by the beamsplitter. The cameras are aligned and calibrated to ensure their images match correctly. See the \supp{} for additional details.
    }
    \label{fig:capture_setup}
\end{figure}

For real-world experimental assessment, we built a prototype to validate our collaborative metalens-based camera design in indoor and outdoor settings (Fig.~\ref{fig:capture_setup}). The setup consists of two cameras: a metalens array camera and a reference camera for comparison. The metalens camera used a CMOS sensor (Allied Vision, Prosilica GT1930C) with \SI{5.86}{\micro\metre} pixel pitch and a metalens array of $2\times3$ lenses with \SI{3.6}{\milli\metre} focal length, \SI{3.57}{\milli\metre} lens pitch and \SI{1.5}{\milli\metre} lens diameter. \rev{The fabrication details of our metasurface optic can be found in \supp{}.} We aligned the metalens array parallel to the sensor using a three-axis translation stage and a two-axis rotatable mount. Each metalens produces a subimage of approximately $\SI{3.5}{\milli\metre}\times\SI{3.5}{\milli\metre}$, covering a FOV of $\SI{50}{\degree}\times\SI{50}{\degree}$. We used a custom optical baffle to prevent crosstalk and added a UV/IR cut-off filter (Edmund Optics, 54-749) to block the invisible spectrum. For reference comparison, we used a separate camera using a FLIR GS3-U3-32S4C-C sensor with a \SI{3.5}{\milli\metre} C-series fixed focal length camera that has a physical length of $\sim$\SI{50}{\milli\metre} from front to back. A 60T/40R beamsplitter (Edmund Optics, 72-502) enabled simultaneous capture through both cameras. We developed a custom graphical user interface for visualization and synchronized camera control with adjustable exposure time and white balance. We captured 653 indoor and outdoor scenes, aligning the reference and metalens images through homography calibration using a ChArUco board pattern. See \supp{} for more details. Notably, our proposed reconstruction method does \textit{not} rely on large-scale training using experimentally captured datasets. Instead, it leverages a much larger amount of synthetic data generated from the rendered hyperspectral parallax dataset described earlier.

\section{Synthetic Assessment}
\label{sec:assessment}

    

\subsection{Broadband PSF Analysis}
\label{sec:aberration_tensor}

\begin{figure}[t]
    \vspace*{-6pt}
    \centering
    \includegraphics[width=1.0\columnwidth]{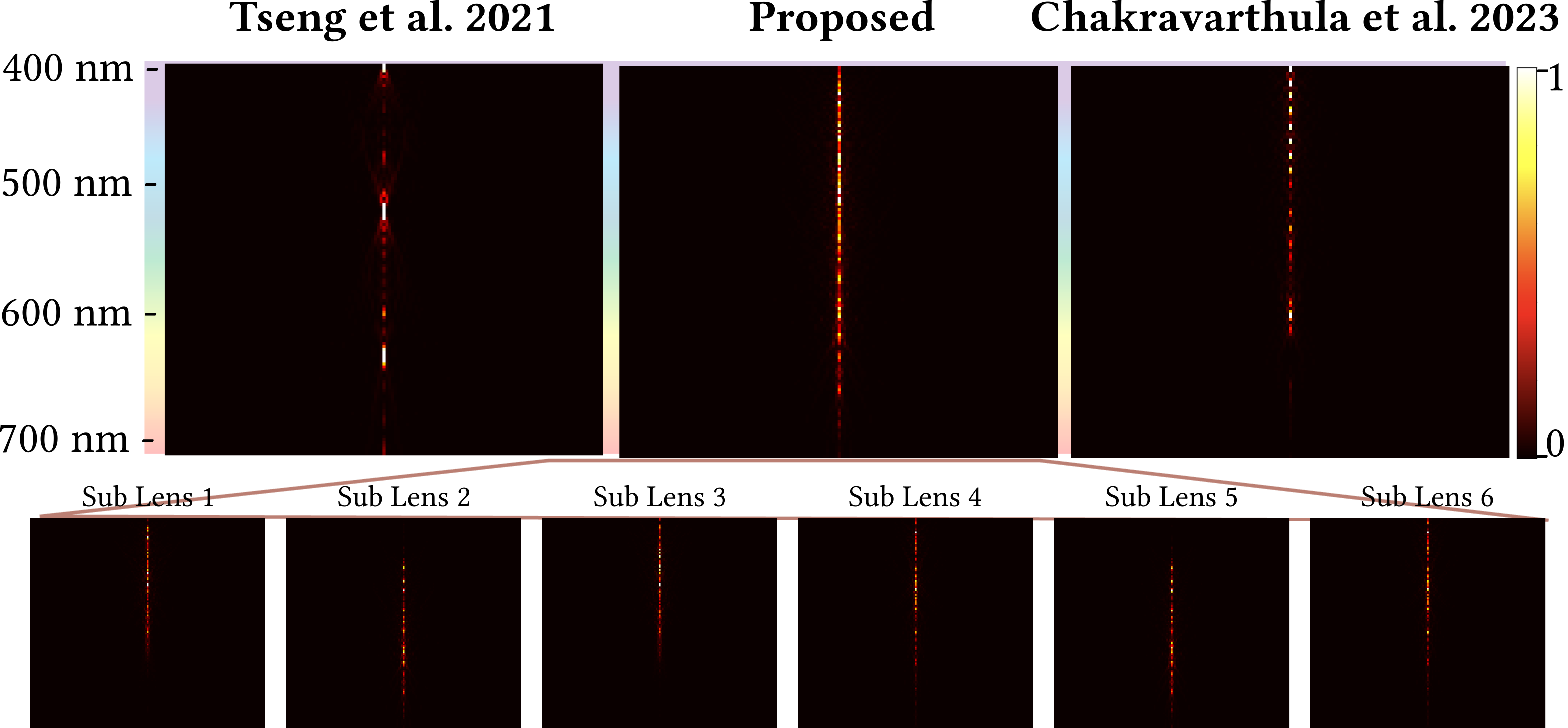}\vspace*{-6pt}
    \caption{\textbf{Analysis of PSFs for existing metalens array imagers.} We plot here X-Z projection of 3D stacked PSFs spanning \SI{400}{\nano\metre} to \SI{700}{\nano\metre} in \SI{1}{\nano\metre} intervals. This visualization highlights the discretely sharp PSFs produced by \cite{neural_nanooptics, chakravarthula2023thin} under broadband illumination, whereas our proposed design, with collaboratively optimized sublenses, maintains a continuously sharp PSF across the entire visible range. We note that the individual sublenses collaborate here to focus over the entire spectrum.}    
    \label{fig:psf_anlysis}
\end{figure}

We start the proposed metalens array evaluation by spectrally resolved PSF analysis.
Specifically, we sample every \SI{1}{\nano\metre} between \SI{400}{\nano\metre} and \SI{700}{\nano\metre} as the incident plane-wave wavelength, \(\Lambda_{\text{disc}}\). For each sampled wavelength and each of the \(K\) lenses, we compute the field energy distribution on the image plane using the angular-dependent PSF model from Section~\ref{sec:formationmodel}, with spatial sampling matching the camera sensor model in Section~\ref{sec:experiment}. The result is a 4D PSF tensor \(P_{\text{disc}}\) of shape \((K,|\Lambda_{\text{disc}}|,N_x, N_y)\), where \(N_x\) and \(N_y\) represent the sensor pixel dimensions behind each individual lens. To visualize this high-dimensional PSF tensor, we take a radial slice $P_{\text{slice}}$, that is
\[
P_{\text{slice}}[:, :,\tfrac{N_x}{2}, :]
\]
and display each sublens PSF slice at the bottom of Fig.~\ref{fig:psf_anlysis} after a global normalization (preserving their relative energy). For an ideal lens, this response for the full visible spectrum is a continuous centered line. We observe that the proposed joint optimization promotes a lens array in which the collective output, combined through the proposed joint reconstruction, approximates an ideal broadband lens. The sub-lenses collaborate to capture the entire broadband spectrum. The top-center ``synthetic effective lens'' is the weighted sum of each sublens contribution, based on their signal-to-noise ratio (SNR) as described in Eq.~\eqref{eq:deconvfilter}. This plot confirms the collaborative nature of the learned array elements. 

We compare to existing broadband metalens imaging techniques \cite{neural_nanooptics,chakravarthula2023thin} with the same analysis as described above. Figure~\ref{fig:psf_anlysis} confirms that these imagers produce only small-footprint PSFs at the discrete wavelengths for which they were optimized (\SI{462}{nm}, \SI{511}{nm}, and \SI{606}{nm} in~\cite{neural_nanooptics}, and \SIrange{400}{700}{nm} in \SI{10}{nm} increments for~\cite{chakravarthula2023thin}). In contrast, our proposed design—with collaboratively optimized sublenses—maintains continuously sharp PSFs across the full visible spectrum. This outcome further highlights the importance of continuous wavelength sampling and collaborative optimization for achieving high-quality broadband metalens imaging.

\subsection{Synthetic Evaluation}

\begin{figure*}[ht!]
\vspace{-14pt}
    \small
    {\sffamily
    \setlength{\tabcolsep}{0pt}
    \renewcommand{\arraystretch}{1}
    \begin{tabular}{C{0.2\textwidth}C{0.2\textwidth}C{0.2\textwidth}C{0.2\textwidth}C{0.2\textwidth}}
    \multicolumn{5}{c}{\includegraphics[width=\linewidth]{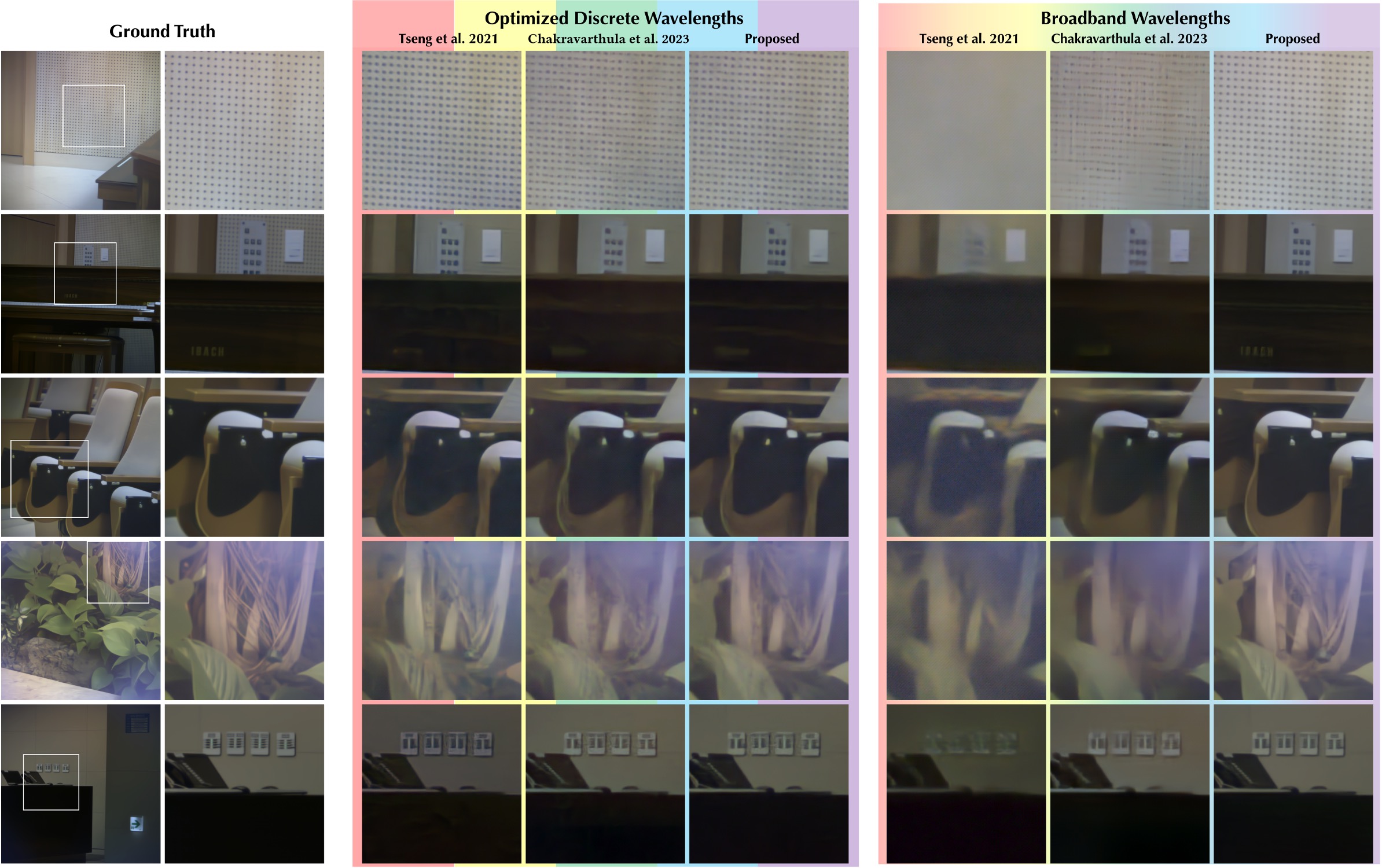}}\\[-12pt]
    \end{tabular}
    }
    \caption{\textbf{Synthetic broadband image reconstruction with flat metasurface cameras.} Due to computational and memory complexity, existing flat metalens optics~\cite{neural_nanooptics, chakravarthula2023thin} have been optimized for discrete wavelength sets. As such, these methods perform comparably to our proposed design \emph{at the discrete spectral bands for which they were optimized for} (\SI{462}{nm}, \SI{511}{nm}, and \SI{606}{nm} for \cite{neural_nanooptics}, and \SI{400}{nm}:\SI{10}{nm}:\SI{700}{nm} for \cite{chakravarthula2023thin}). However, their quality deteriorates significantly under real-world continuous broadband illumination (\SI{400}{nm}:\SI{1}{nm}:\SI{700}{nm}). In contrast, the proposed design method, along with the array design, lifts wavelength sampling limitations, and the full-spectrum collaborative array consistently delivers superior image quality across the full spectrum.}
    \label{fig:discrete_continous_comparison}
\end{figure*}

\paragraph{Evaluation of Broadband Performance}


We confirm the trend from the broadband PSF analysis above for image reconstruction. To this end, we simulate measurements using a real-world hyperspectral dataset~\cite{jeon2024spectral} as input. Specifically, the sensor measurement is generated by following the noise model in Section~\ref{sec:formationmodel} with a Poissonian-Gaussian noise model. To handle any spectral mismatch between the sampled wavelength and dataset wavelength, we interpolate the input hyperspectral images. We then convert the hyperspectral ground truth image, metalens measurement, and PSF back into the RGB domain according to the camera’s spectral response function. The same joint deconvolution algorithm from Section~\ref{sec:alignment} is applied for reconstruction except for the parallax alignment module. 

First, we evaluate performance at the \emph{discrete design wavelengths} for which other respective optics are optimized for: (\SI{462}{nm}, \SI{511}{nm}, and \SI{606}{nm} in~\cite{neural_nanooptics}, and \SIrange{400}{700}{nm} in \SI{10}{nm} increments for \cite{chakravarthula2023thin}). To maintain consistency, we test our proposed array using the same sampling grid of \([\SI{400}{nm}:\SI{10}{nm}:\SI{700}{nm}]\), even though it was \emph{not explicitly optimized for this discrete wavelength set}. As reported in Fig.~\ref{fig:discrete_continous_comparison}, all three imagers yield comparable reconstruction quality, apart from minor color deviations in~\cite{neural_nanooptics}, which uses fewer sampling wavelengths. This result validates the effectiveness of the alternate designs for their intended discrete-wavelength range.

Next, we test each imager under \emph{continuous broadband illumination}, for which we approximate with discrete 400 to 700 nm wavelength sampling with 1 nm spacing. In agreement with the broadband PSF analysis, the two baseline methods experience a notable drop in performance, owing to the poor SNR at the unoptimized bands. In contrast, our collaborative design performs {robustly across the continuous broadband illumination, where the other methods severely degrade}. Notably, our design even performs better under broadband conditions than under discrete-wavelength illumination, because the PSFs used for deconvolution align more closely with the physical system, and the overall quality is higher when more spectral bands contribute to the measurement.

\begin{table}[!t]
\centering
\caption{\textbf{Quantitative evaluation of broadband image reconstruction with flat metasurface imagers.} We report here quantitative measurements corresponding to the experiments visualized in Fig.~\ref{fig:discrete_continous_comparison}. 
The evaluations confirm the qualitative trends. While existing metalens imaging methods suffer from lower SNR across unoptimized bands, our collaborative design remains robust for broadband imaging, see text.} 
\vspace{-2mm}
\begin{tabular}{lccc}
\toprule
& \small Tseng & \small Chakravarthula & \\
& \small et al. \shortcite{neural_nanooptics} \hspace{8pt} & \small et al. \shortcite{chakravarthula2023thin} \hspace{8pt} & \small Proposed \\
\midrule
SSIM $\uparrow$ & 0.72 & 0.80 & \textbf{0.95} \\
PSNR [dB] $\uparrow$ & 25.92 & 27.34 & \textbf{34.03} \\
1-LPIPS $\uparrow$ & 0.55 & 0.62 & \textbf{0.87} \\
\bottomrule
\end{tabular}
\label{tab:broadband}
\end{table}

\paragraph{Validation of Collaboration}

In addition to the broadband PSF results shown in Fig.~\ref{fig:psf_anlysis}, which indicate that individual sublenses collaboratively exhibit small-footprint PSFs, we also optimized a single broadband metalens using the same methodology described in Section~\ref{sec:Imaging} to compare with the proposed design for validating the effectiveness of the collaborative array design. 

\begin{figure}[t]
    \vspace*{-6pt}
    \centering
    \includegraphics[width=1.0\columnwidth]{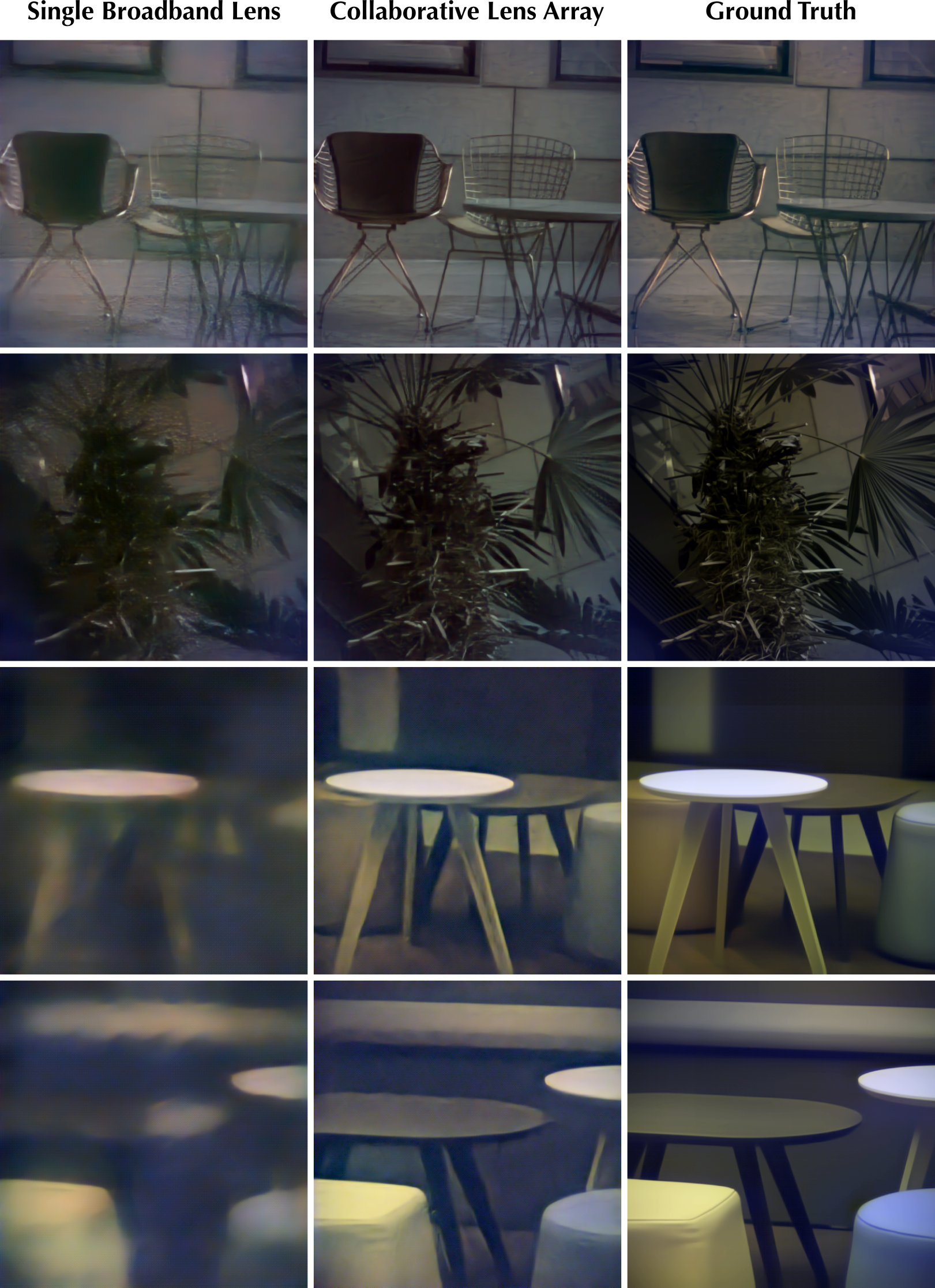}\vspace*{-6pt}
    \caption{\textbf{Comparison between a single broadband-optimized lens and the proposed collaborative array under challenging illumination spectra.} Although a single metalens can be continuously optimized for broadband illumination within our framework (left), a single lens struggles to focus the entire visible spectrum. In contrast, our method employs multiple collaboratively optimized lenses to overcome these limitations. We confirm this here qualitatively with reconstructions under challenging illumination scenarios—such as heavily tilted spectra or discrete illumination~\cite{li2021multispectral}—demonstrating that our collaborative array design consistently outperforms the single broadband lens across a variety of illumination conditions, see text.}
    \label{fig:single_multi_comparison}
\end{figure}

To simulate realistic real-world captures under challenging illumination conditions, we utilize the hyperspectral illuminator dataset from~\cite{li2021multispectral} in combination with the hyperspectral imaging dataset mentioned in~\cite{jeon2024spectral}. Specifically, we select 10 illuminators with tilted or discrete spectral characteristics and uniformly sample wavelengths in the range \([\SI{400}{nm}:\SI{10}{nm}:\SI{700}{nm}]\). Sensor measurements are then generated by computing a hyperspectral illumination weighted sum of the hyperspectral measurements, which are obtained via hyperspectral domain convolution between the hyperspectral data and the PSFs. Representative reconstruction outputs are shown in Fig.~\ref{fig:single_multi_comparison}.

Although both imagers face substantial challenges due to the extreme illumination conditions, the single broadband-optimized lens shows noticeably more variation in SNR across different illuminators. In contrast, the proposed lens array design remains robust, producing consistently high-fidelity reconstructions. This underscores the spectral bandwidth limitations \cite{Presutti2020FocusingLimits} of a single metalens in maintaining strong broadband performance while highlighting the robustness of our collaborative array approach.

\paragraph{Ablation Experiments}

\begin{figure*}[ht!]
    \vspace{-4pt}
    \small
    {\sffamily
    \setlength{\tabcolsep}{0pt}
    \renewcommand{\arraystretch}{1}
    \begin{tabular}{C{0.2\textwidth}C{0.2\textwidth}C{0.2\textwidth}C{0.2\textwidth}C{0.2\textwidth}}
    \multicolumn{5}{c}{\includegraphics[width=\linewidth]{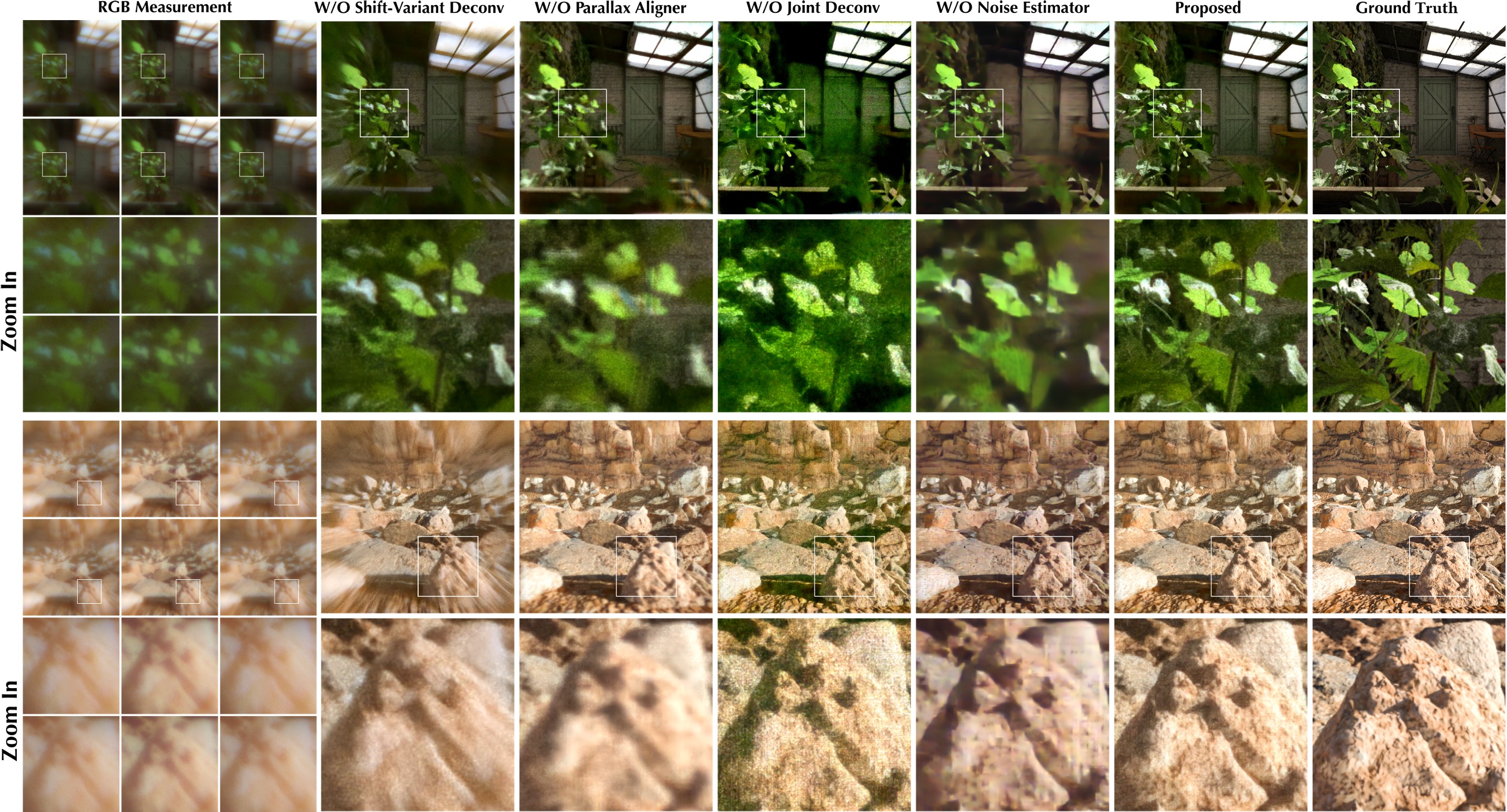}}\\[-12pt]
    \end{tabular}
    }
    \caption{\textbf{Qualitative ablation experiments.} We evaluate our reconstruction pipeline by comparing three ablated variants on the hyperspectral parallax-aware metalens array dataset: 
    (i)~replacement of the spatial-variant deconvolution with spatial-invariant deconvolution using on-axis PSF,
    (ii)~removal of the parallax-aligner, 
    (iii)~replacement of the joint deconvolution step with independent Wiener deconvolution, and 
    (iv)~removal of the noise neural estimator. 
    Using spatial-invariant deconvolution instead of spatial-variant deconvolution results in astigmatism at the periphery of the FOV. Methods without the parallax aligner exhibit blurry reconstructions in regions with closer object distance due to misalignments. 
    Replacing the joint deconvolution with independent deconvolution leads to pronounced color aberrations, as individual metalenses are not broadband. 
    Finally, omitting the noise neural estimator hampers noise removal under the metalens array’s broadband varying SNR for each scene wavelength, compromising detail preservation.}
    \label{fig:ablation_study}
\end{figure*}

To validate the effectiveness and all design choices of our proposed reconstruction method, we conduct ablation experiments on the synthetic hyperspectral parallax dataset introduced in Section~\ref{subsec:hs_parallax_dataset}. We compare five variants of our method:
(i)~the pipeline without shift-variant deconvolution,
(ii)~the pipeline without the parallax aligner, 
(iii)~the pipeline without the joint deconvolution module,
(iv)~the pipeline without the noise estimator, and
(v)~the full pipeline (proposed). Qualitative and quantitative results are reported in Fig.~\ref{fig:ablation_study} and Table~\ref{tab:ablation}.

All ablated methods operate with the same array measurements and PSFs. Specifically, we randomly sample 10 distinct wavelengths from the visible spectrum and generate the corresponding hyperspectral ground-truth image using the hyperspectral lifting method provided by our rendered dataset. We then compute the hyperspectral PSFs of the metalens array for each sampled wavelength. The RGB lens-array measurement with parallax is synthesized by convolving the ground-truth image patches with the wavelength-specific PSFs that correspond to the incident angle of the patches, integrating the results according to the CSRF, and finally adding Poissonian-Gaussian noise~\cite{Foi2008PracticalPN} with parameters $(a,b) = (4\times10^{-5}, 1 \times 10^{-6})$ estimated from experimental raw captures. 

\begin{table}[t!]
    \setlength{\tabcolsep}{0em}
    \centering
    \footnotesize
    \caption{
        \textbf{Quantitative ablation study.} We simulate RGB shift-variant array measurements using a hyperspectral parallax dataset and corresponding PSFs to assess our proposed reconstruction method. Each sub-module (spatial-variant deconvolution, parallax aligner, joint deconvolution, and SNR estimator) is removed in turn, and the resulting models are evaluated on SSIM, PSNR, and LPIPS~\shortcite{zhang2018unreasonable}. Our findings indicate that each sub-module serves a distinct yet essential function in the final reconstruction. By combining all sub-modules, the full reconstruction method yields the best overall performance.
        }
    \begin{tabularx}{\linewidth}{m{0.15\linewidth}XXXXXX}
    \toprule
    &
    {\footnotesize W/O SV Deconv } &
    {\footnotesize W/O Parallax Aligner} &
    {\footnotesize W/O Joint Deconv} &
    {\footnotesize W/O SNR Estimator} &
    {\footnotesize Proposed} \\
    \midrule
    SSIM $\uparrow$ & 
    0.53 & 
    0.62 & 
    0.37 & 
    0.61 & 
    \textbf{0.67} \\
    PSNR [dB] $\uparrow$ & 
    18.32 & 
    19.74 & 
    18.90 & 
    20.18 & 
    \textbf{21.14} \\

    1-LPIPS $\uparrow$ & 
    0.41 & 
    0.49 & 
    0.35 & 
    0.46 & 
    \textbf{0.52} \\
    \bottomrule
\end{tabularx}
    
    \label{tab:ablation}
    \vspace{-5mm}
\end{table}

\medskip
\noindent
\textbf{Without Shift-Variant Deconvolution.}
In our short focus design, the large incident angles at the periphery of the FOV induce astigmatism. When the system’s shift-variance is ignored -- using only the on-axis PSF for deconvolution -- resolution degrades and color artifacts appear, as shown in Fig.~\ref{fig:ablation_study}. By contrast, applying patch-wise shift-variant deconvolution accounts for off-axis aberrations and effectively corrects the degradation, resulting in improved image quality.

\medskip
\noindent
\textbf{Without Parallax Aligner.}
Removing the neural warping operations for parallax correction degrades reconstruction quality, especially for objects closer to the camera where misalignment becomes more pronounced. Although the baseline distance between the individual metalenses is relatively small, reconstruction on our 3D rendered parallax dataset shows a noticeable drop in reconstruction quality for objects located within one meter of the camera.

\medskip
\noindent
\textbf{Without Joint Deconvolution.}
Next, we investigate replacing the joint deconvolution module with six independent deconvolution branches. The resulting per-metalens reconstructions are then averaged to form the input to the remaining pipeline. Since individual metalenses cannot maintain a consistently high SNR across a broad range of wavelengths, this independent approach produces chromatic artifacts. Furthermore, averaging these color-degraded images only compounds the blur and color inaccuracies.

\medskip
\noindent
\textbf{Without Noise Estimator.}
Finally, we remove the neural noise estimator and retrain the network while keeping the parallax aligner and joint deconvolution module weights fixed. Due to the wavelength-dependent SNR response of the metalens, the output from the deconvolution module exhibits varying degrees of colored noise amplification based on the MTF under different sampled wavelengths. As a result, the network without noise estimator struggles to disentangle the signal from the various noise level measurements, leading to degraded reconstructions~\cite{guo2019toward}.

Overall, the ablation experiments from above confirm the effectiveness of each component in the proposed reconstruction method. 

\begin{figure*}[p!]
\vspace{-18pt}
    \small
    {\sffamily
    \setlength{\tabcolsep}{0pt}
    \renewcommand{\arraystretch}{1}
    \begin{tabular}{C{0.2\textwidth}C{0.2\textwidth}C{0.2\textwidth}C{0.2\textwidth}C{0.2\textwidth}}
    \multicolumn{5}{c}{\includegraphics[width=\linewidth]{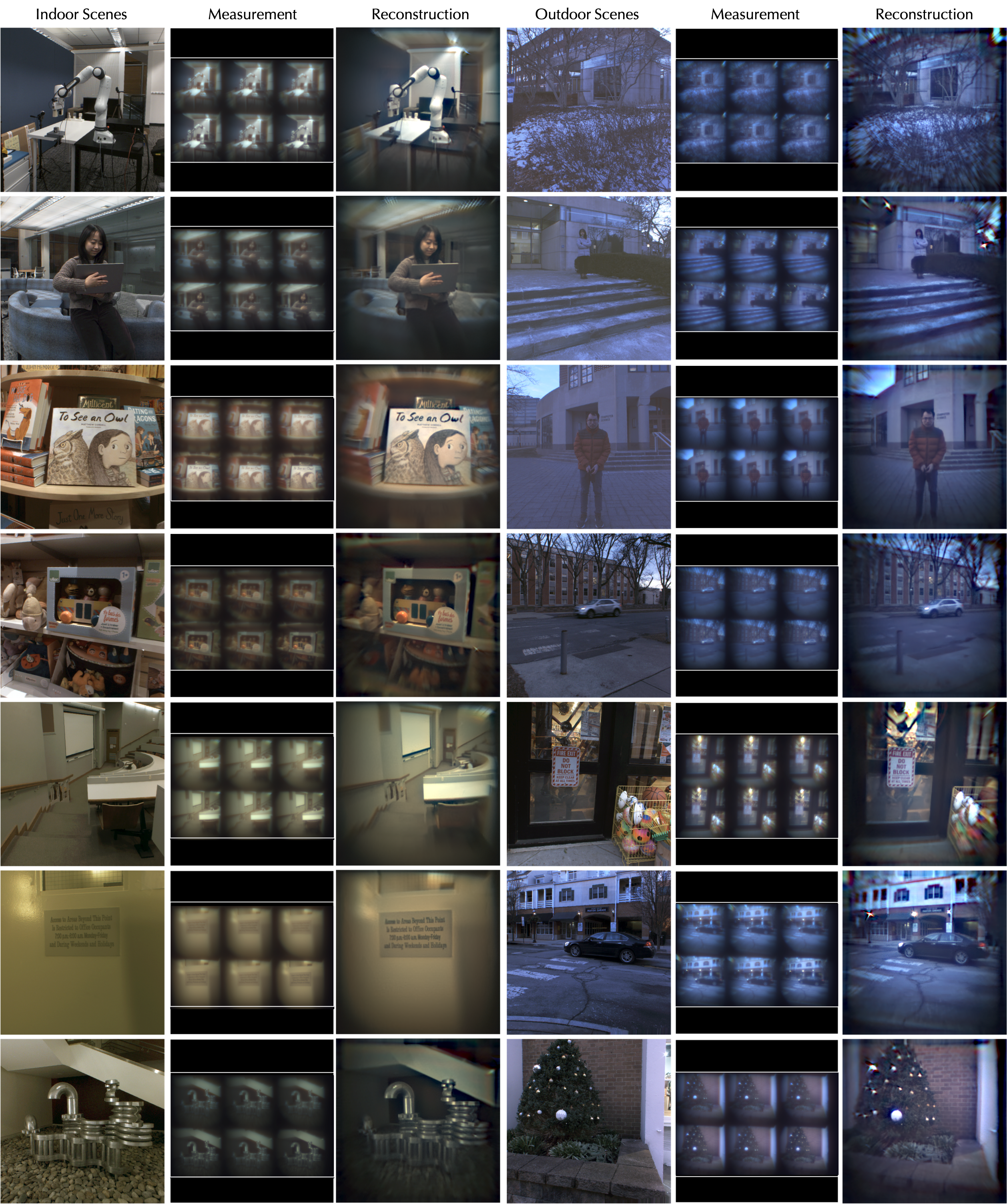}}\\[-12pt]
    \end{tabular}
    }
    \caption{\textbf{Experimental reconstructions of indoor and outdoor scenes.} The broadband-optimized metalens design yields a sharp array of image measurements with small chromatic aberration across diverse illumination conditions. The reconstructed image quality and color fidelity approach that of the reference images, captured by a conventional multi-element camera at a comparable F-number.}
    \label{fig:experimental_captures}
\end{figure*}

\begin{figure*}[p!]
\vspace{-18pt}
    \small
    {\sffamily
    \setlength{\tabcolsep}{0pt}
    \renewcommand{\arraystretch}{1}
    \begin{tabular}{C{0.2\textwidth}C{0.2\textwidth}C{0.2\textwidth}C{0.2\textwidth}C{0.2\textwidth}}
    \multicolumn{5}{c}{\includegraphics[width=\linewidth]{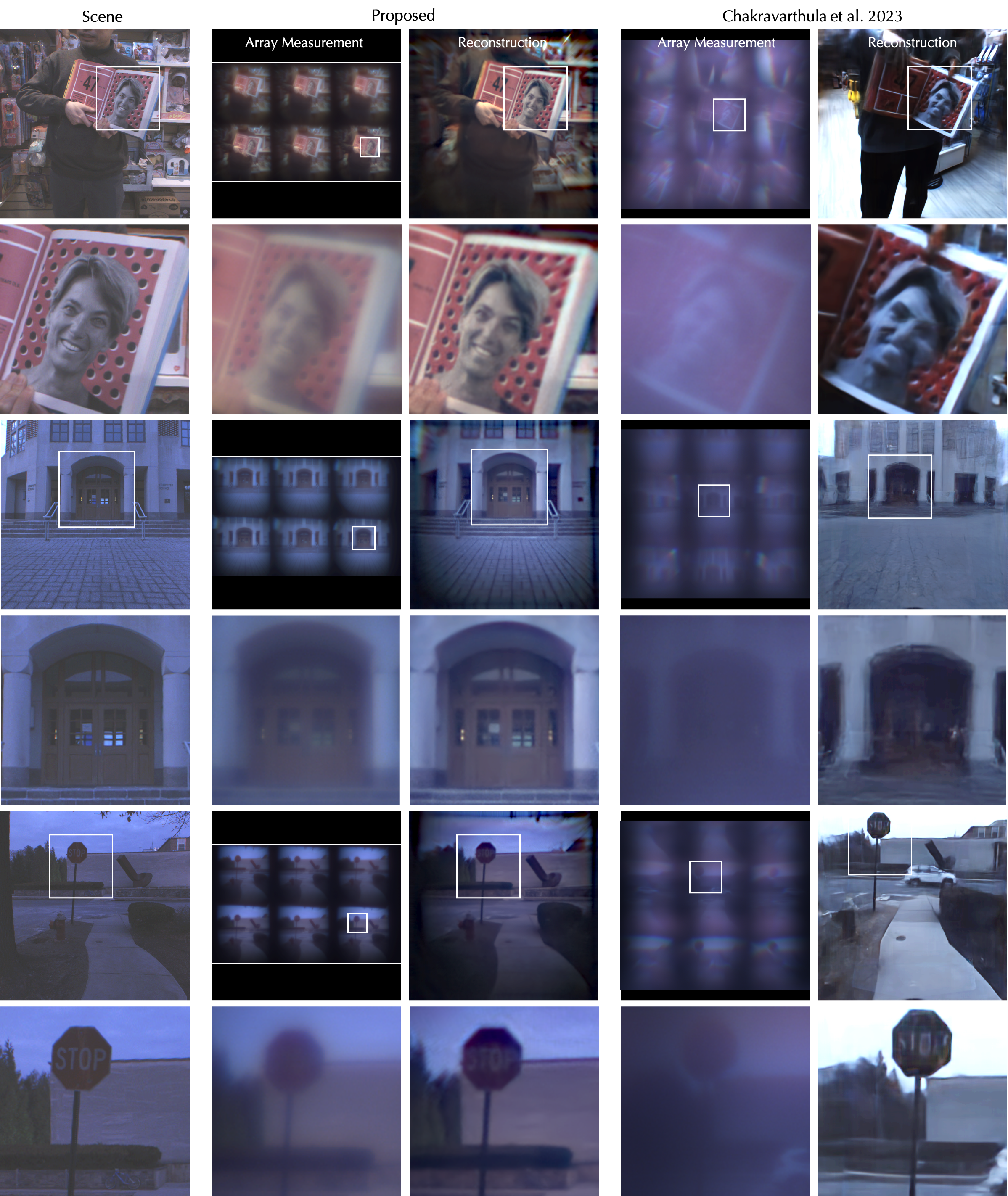}}\\[-12pt]
    \end{tabular}
    }
    \caption{\textbf{Experimental comparison to~\cite{chakravarthula2023thin}.} For similar scenes and illumination conditions, the proposed collaborative metalens array provides more accurate measurements and image reconstructions than~\cite{chakravarthula2023thin}. The image zoom-ins show the proposed collaborative metalens array can capture a more broadband focused raw measurement, and a more plausible reconstruction of the scene, revealing details such as the portrait, building door windows, and a stop sign.}
    \label{fig:nano_array_comparison}
\end{figure*}

\section{Experimental Assessment}

\label{subsec:real_world_capture}

To validate the proposed design experimentally across diverse and challenging conditions---including variations in illumination, depth parallax, and the presence of novel objects---we conducted real-world experiments using our fabricated metasurface. Using the prototype camera design described in Section~\ref{subsec:prototype}, we captured scenes featuring small text, human faces, significant depth variations, and low-contrast textures. These scenes were recorded under a wide range of illumination conditions such as low-light, artificial light, direct sunlight, and intense localized light sources. \rev{Typical exposure times were \SI{100}{\milli\second} for indoor and \SI{30}{\milli\second} for outdoor scenes, and could be reduced by removing the 60T/40R beam-splitter.} Representative measurement and reconstruction results are shown in Fig.~\ref{fig:experimental_captures}, alongside aligned comparison measurements from a reference \SI{3.5}{\milli\metre} C-series fixed focal length camera that has a physical length of $\sim$\SI{50}{\milli\metre}.
Each metalens in the array captures broadband scene measurements of objects in both indoor and outdoor scenes at the center of each metalens. \rev{The dark bands separating adjacent sub‐images arise from the integrated optical baffle, which prevents stray light and suppresses crosstalk between neighboring apertures.} \rev{Our reconstruction method mitigates the spatially-varying optical aberrations, improving the image contrast in the center and the periphery with finer details.}
See \supp{} for more results, and the \emph{Supplementary Video} for dynamic scenes.


\paragraph{Qualitative Comparison to Existing Array Cameras}
Next, in Fig.~\ref{fig:nano_array_comparison}, we compare our camera prototype with the metalens array camera proposed in~\cite{chakravarthula2023thin}, whose
image reconstruction technique is based on a diffusion prior.
The proposed method achieves improved detail and substantially less scattering in the raw measurement: for instance, detail on the human face, the building entrance or the traffic sign in Fig.~\ref{fig:nano_array_comparison}. These raw measurements validate densely sampling of the wavelengths and incident angles
during optical optimization, as well as the accuracy of the proposed neural proxy modeling in Section~\ref{sec:formationmodel}.

We surpass the diffusion-based approach of \cite{chakravarthula2023thin} both in terms of accuracy and absence of `hallucinated content' in the reconstructed results. We correctly recover details such as
the face of the person and the polka dots on the book, the windows on the building entrance, or even the text
`stop' on the traffic sign, whereas the compared method instead smooths the face and mangles
the dots on the book, and introduces visual artifacts on the entrance and the stop sign
that is hardly readable.
This is due to the generative prior of~\cite{chakravarthula2023thin} trained to lift a {\em single} low-contrast measurement to a sharper full-color result.
%
In contrast to the proposed method, hallucination appears in the prediction of details where there is an ambiguity about the underlying sharp image. This is illustrated in Fig.~\ref{fig:teaser} where the generative model has restored `65' whereas the original number is `86'.


\paragraph{Runtime Performance}
Our reconstruction runs at real-time rates. Compared to the diffusion module of \cite{chakravarthula2023thin} that takes more than a minute to process a single image, our technique is $2000\times$ faster and runs at 35 FPS on the same test platform. Both approaches are implemented in PyTorch and do not leverage further optimizations. We rely on fast FFT-based inverse filtering with a subsequent denoising network that is lightweight since fusing six images already reduces noise\footnote{If we assume the noise to be purely Gaussian, merging 6 images corresponds to a $\sqrt{6}\approx 2.45$ noise reduction factor.}. The runtime bottleneck stems from the dense optical flow prediction with the optical flow model~\cite{teed20raft} (which takes up about 60\% of the time for reconstructing a frame with our technique).

\section{Discussion}
\label{sec:limitation}





\rev{Our collaborative metalens-array prototype delivers high-fidelity, broadband imaging under diverse illumination but it remains a research demonstrator rather than a product-grade camera due to its form factor. To facilitate prototyping, we retained the Sony IMX174 sensor cover glass and mounted a 7 mm × 10.5 mm metalens array just 1 mm above it. The array dimensions match the sensor 7.1 mm × 11 mm active area, and the resulting optics yield a 3.6 mm focal length. In an industrialized, smartphone-style implementation—with integrated inner baffles and f/2 optics on a 12 MP sensor—the array could shrink to a 2 × 2 layout of 1.5 mm elements, occupying only a 5 mm × 5 mm footprint at a 3 mm standoff while delivering a 60° field of view. This ultra-thin configuration aligns with smartphone packaging constraints—minimizing optical thickness while efficiently utilizing the PCB footprint, whereas comparable refractive solutions would either degrade into a low-quality fisheye or require an impractical, multi-element assembly. Despite the throughput trade-offs inherent to high-NA metaoptics, our prototype still supports real-time, video-rate capture.}

\rev{Our reconstruction pipeline runs at 35 FPS on laptop‐class GPUs and stands to gain further image‐quality improvements from ring deconvolution \cite{kohli2022ring} and sub‐aperture super‐resolution \cite{Eboli2022fast}. Pruning and fine-tuning a lightweight optical-flow network, denoiser distillation, and dedicated ISP hardware could further accelerate processing for potential on-edge deployment. The array's angular multiplexing also enables direct depth estimation from raw captures \cite{venkataraman2013picam}. At the design level, our optimization framework supports full-angle, multi-spectral phase-response simulations for arbitrary metalens array geometries free from radial-symmetry constraints at the cost of memory overhead. By restricting the phase profile to radial symmetry under on-axis illumination, one can instead take advantage of a one-dimensional Rayleigh-Sommerfeld diffraction model to cut memory requirements~\cite{froch2025beating}, trading off large-angle performance. Finally, extending our neural surrogate to capture angle- and wavelength-dependent transmission~\cite{hazineh2022d, wirth2025wide} could further enhance design fidelity and unlock new high-NA meta-optic applications.}
\section{Conclusion}
\label{sec:conclusion}

We have introduced a novel computational imager using collaborative metasurface arrays that enables broadband imaging in an ultra-thin form factor. Through the integration of a learned neural proxy for nanostructure-to-phase mapping, jointly designed collaborative reconstruction, and large-scale distributed optimization of 100 million \rev{nanoposts}, our design framework achieves significant improvements to prior metalens array designs in both raw and reconstruction image quality. We validate the method with an experimental prototype that successfully captures and reconstructs broadband images in both indoor and outdoor scenes under varying illumination spectra. 

In the future, one exciting avenue for research is to correct spatially-varying aberrations, which are an intrinsic difficulty of single-layer optics, through multiple metasurface elements in a thin stack, with the expanded parameter space being manageable by our framework. Looking forward, we hope that our method for ultra-thin metasurface imagers serves as a systematic tool for exploring new large-area optics applications, including large-scale manufacturing of on-chip metasurface imaging sensors, collaborative optimization for polarization-sensitive and hyperspectral metasurface imaging, and specialized meta-imaging systems. With the broadband capabilities enabled by this work, we hope that on-chip imaging sensors realize the potential to enable broad applications across the domains -- from tiny camera arrays in micro-robotics to wearable flat cameras in healthcare.

\begin{acks}
We thank Dongyu Du, Jiwoon Yeom, Zheng Shi, and Johannes Fröech for their fruitful discussion on the project. Felix Heide gratefully acknowledges support from an NSF CAREER Award (2047359), a Packard Foundation Fellowship, a Sloan Research Fellowship, a Disney Research Award, a Sony Young Faculty Award, a Project X Innovation Award, and an Amazon Science Research Award. Wolfgang Heidrich acknowledges support from KAUST Individual Baseline Funding. Fabrication was carried out at the Washington Nanofabrication Facility/Molecular Analysis Facility at the University of Washington—a National Nanotechnology Coordinated Infrastructure site supported by the National Science Foundation under Grant Nos. NSF-2127235 and NSF-2223495 (EFRI-BRAID).
\end{acks}



\bibliographystyle{ACM-Reference-Format}
\bibliography{references}
\end{document}